\documentclass[useAMS]{mn2e}
\usepackage{natbib}
\usepackage{url,times,graphicx,amsmath,amsfonts,amssymb,aas_macros,epsfig,epstopdf}
\usepackage{multicol}
\usepackage[
    pdftex,
    a4paper=true,
    breaklinks=true,
    bookmarks=true,
    bookmarksopen=false,
    bookmarksopenlevel=2
    bookmarksnumbered=true,
    bookmarkstype=toc,
    colorlinks=true,
    citecolor=blue,
    linkcolor=blue,
    menucolor=green,
    urlcolor=magenta,
]{hyperref}

\newcommand{\art}{\textsc{ART}}
\newcommand{\arepo}{\textsc{Arepo}}
\newcommand{\hydra}{\textsc{Hydra}}

\newcommand{\gadget}{\textsc{Gadget}}
\newcommand{\gadgetx}{\textsc{Gadget3-X}}
\newcommand{\gadgetmagneticum}{\textsc{Gadget3-Magneticum}}
\newcommand{\gadgettwox}{\textsc{Gadget2-X}}
\newcommand{\gadgetowls}{\textsc{Gadget3-OWLS}}

\newcommand{\gadgetmusic}{\textsc{Gadget3-MUSIC}}
\newcommand{\gadgetsphs}{\textsc{Gadget3-SPHS}}
\newcommand{\gadgetwindy}{\textsc{Gadget3-PESPH}}
\newcommand{\gadgetanarchy}{\textsc{Gadget2-Anarchy}}
\newcommand{\ramses}{\textsc{RAMSES}}
\newcommand{\enzo}{\textsc{ENZO}}

\newcommand{\hMpc}{{\ifmmode{h^{-1}{\rm Mpc}}\else{$h^{-1}$Mpc}\fi}}
\newcommand{\hkpc}{{\ifmmode{h^{-1}{\rm kpc}}\else{$h^{-1}$kpc}\fi}}
\newcommand{\hMsun}{{\ifmmode{h^{-1}{\rm {M_{\odot}}}}\else{$h^{-1}{\rm{M_{\odot}}}$}\fi}}
\newcommand{\Mstar}{{\ifmmode{M_{*}}\else{$M_{*}$}\fi}}
\newcommand{\Mhalo}{{\ifmmode{M_{\rm halo}}\else{$M_{\rm halo}$}\fi}}
\newcommand{\Ngal}{{\ifmmode{N_{\rm gal}}\else{$N_{\rm gal}$}\fi}}
\newcommand{\Norph}{{\ifmmode{N_{\rm orphan}}\else{$N_{\rm orphan}$}\fi}}
\newcommand{\Nxorph}{{\ifmmode{N_{\rm non-orphan}}\else{$N_{\rm non-orphan}$}\fi}}
\newcommand{\Zsolar}{{\ifmmode{Z_{\odot}}\else{$Z_{\odot}$}\fi}}
\newcommand{\ltsima}{$\; \buildrel < \over \sim \;$}
\newcommand{\gtsima}{$\; \buildrel > \over \sim \;$}
\newcommand{\lsim}{\lower.5ex\hbox{\ltsima}}
\newcommand{\gsim}{\lower.5ex\hbox{\gtsima}}

\def\lesssim{\mathrel{\hbox{\rlap{\hbox{\lower4pt\hbox{$\sim$}}}\hbox{$<$}}}}
\def\gtrsim{\mathrel{\hbox{\rlap{\hbox{\lower4pt\hbox{$\sim$}}}\hbox{$>$}}}}

\newcommand{\beq}{\begin{equation}}
\newcommand{\eeq}{\end{equation}}
\def\beqa{\begin{eqnarray}}
\def\eeqa{\end{eqnarray}}
\def\hMpc{$h^{-1}\,{\rm Mpc}$}
\def\hkpc{$h^{-1}\,{\rm kpc}$}

\def\Gadget{{\sc Gadget}}

\def\Hydra{{\sc Hydra}}

\def\Ramses{{\sc Ramses}}

\def\Arepo{{\sc Arepo}}

\begin{document}
\title[nIFTy Cluster Comparison I]{nIFTy galaxy cluster simulations I: dark matter \& non-radiative models}
\author[Sembolini et. al]{Federico Sembolini,$^{1,2,}$\thanks{E-mail: federico.sembolini@uam.es}
Gustavo Yepes,$^{1}$ 
Frazer R. Pearce,$^{3}$ 
Alexander Knebe,$^{1}$ 
\newauthor Scott T. Kay,$^{4}$ 
Chris Power,$^{5}$ 
Weiguang Cui,$^{5}$ 
Alexander M. Beck,$^{6,7,8}$ 
Stefano Borgani,$^{9,10,11}$ 
\newauthor Claudio Dalla Vecchia,$^{12,13}$ 
Romeel Dav\'e,$^{14,15,16}$  
Pascal Jahan Elahi,$^{17}$ 
Sean February,$^{18}$ 
\newauthor Shuiyao Huang$^{27}$
Alex Hobbs,$^{19}$
Neal Katz$^{19}$
Erwin Lau,$^{20,21}$
Ian G. McCarthy,$^{22}$ 
\newauthor Guiseppe Murante,$^{9}$ 
Daisuke Nagai,$^{20,21,23}$ 
Kaylea Nelson,$^{21,23}$
Richard D. A. Newton,$^{5,6}$
\newauthor Ewald Puchwein,$^{24}$
Justin I. Read,$^{25}$ 
Alexandro Saro,$^{14}$ 
Joop Schaye, $^{27}$
Robert J. Thacker$^{28}$
\\
$^{1}$Departamento de F\'isica Te\'{o}rica, M\'{o}dulo 8, Facultad de Ciencias, Universidad Aut\'{o}noma de Madrid, 28049 Madrid, Spain\\
$^{2}$Dipartimento di Fisica, Sapienza Universit$\grave{a}$  di Roma, Piazzale Aldo Moro 5, I-00185 Roma, Italy\\
$^{3}$School of Physics \& Astronomy, University of Nottingham, Nottingham NG7 2RD, UK\\
$^{4}$Jodrell Bank Centre for Astrophysics, School of Physics and Astronomy, The University of Manchester, Manchester M13 9PL, UK\\
$^{5}$International Centre for Radio Astronomy Research, University of Western Australia, 35 Stirling Highway, Crawley, Western Australia 6009, Australia\\
$^{6}$University Observatory Munich, Scheinerstr. 1, D-81679 Munich, Germany\\
$^{7}$Max Planck Institute for Extraterrestrial Physics, Giessenbachstr. 1, D-85748 Garching, Germany\\
$^{8}$Max Planck Institute for Astrophysics, Karl-Schwarzschild-Str. 1, D-85741 Garching, Germany\\
$^{9}$Astronomy Unit, Department of Physics, University of Trieste, via G.B. Tiepolo 11, I-34143 Trieste, Italy\\
$^{10}$INAF - Osservatorio Astronomico di Trieste,  via G.B. Tiepolo 11, I-34143 Trieste, Italy\\
$^{11}$INFN - Sezione di Trieste, via Valerio 2, I-34127 Trieste, Italy\\
$^{12}$Instituto de Astrof\'isica de Canarias, C/ V\'ia L\'actea s/n, 38205 La Laguna, Tenerife, Spain\\
$^{13}$Departamento de Astrof\'sica, Universidad de La Laguna, Av.~del Astrof\'isico Franciso S\'anchez s/n, 38206 La Laguna, Tenerife, Spain\\
$^{14}$Physics Department, University of Western Cape, Bellville, Cape Town 7535, South Africa\\
$^{15}$South African Astronomical Observatory, PO Box 9, Observatory, Cape Town 7935, South Africa\\
$^{16}$ African Institute of Mathematical Sciences, Muizenberg, Cape Town 7945, South Africa\\
$^{17}$Sydney Institute for Astronomy, A28, School of Physics, The University of Sydney, NSW 2006, Australia\\
$^{18}$Center for High Performance Computing, CSIR Campus, 15 Lower Hope Street, Rosebank, Cape Town 7701, South Africa\\
$^{19}$Astronomy Department, University of Massachusetts, Amherst, MA 01003, USA\\
$^{20}$Institute for Astronomy, Department of Physics, ETH Zurich, Wolfgang-Pauli-Strasse 16, CH-8093, Zurich, Switzerland\\
$^{21}$ Department of Physics, Yale University, New Haven, CT 06520, USA\\
$^{22}$Yale Center for Astronomy and Astrophysics, Yale University, New Haven, CT 06520, USA\\
$^{23}$Astrophysics Research Institute, Liverpool John Moores University, 146 Brownlow Hill, Liverpool L3 5RF, UK\\
$^{24}$Department of Astronomy, Yale University, New Haven, CT 06511, USA\\ 
$^{25}$Institute of Astronomy and Kavli Institute for Cosmology, University of Cambridge, Madingley Road, Cambridge CB3 0HA, UK\\
$^{26}$Department of Physics, University of Surrey, Guildford, GU2 7XH, Surrey, United Kingdom\\
$^{27}$Leiden Observatory, Leiden University, P.O. Box 9513, 2300 RA Leiden, the Netherlands\\
$^{28}$Department of Astronomy and Physics, Saint Mary's University, 923 Robie Street, Halifax, Nova Scotia, B3H 3C3, Canada
}
\date{Accepted XXXX . Received XXXX; in original form XXXX}

\pagerange{\pageref{firstpage}--\pageref{lastpage}} \pubyear{2015}
\maketitle

\label{firstpage}

\begin{abstract}
We have simulated the formation of a galaxy cluster in a $\Lambda$CDM
universe using twelve different codes modeling only gravity and non-radiative hydrodynamics (\art, \arepo,
\hydra\ and 9 incarnations of \gadget). This range of codes includes
particle based, moving and fixed mesh codes as well as both Eulerian
and Lagrangian fluid schemes. The various \gadget\ implementations
span traditional and advanced smoothed-particle hydrodynamics (SPH) schemes.  The goal of this
comparison is to assess the reliability of cosmological hydrodynamical
simulations of clusters in the simplest astrophysically relevant case,
that in which the gas is assumed to be non-radiative.  We compare
images of the cluster at $z=0$, global properties such as
mass, and radial profiles of various dynamical and thermodynamical
quantities. The underlying gravitational framework can be aligned very
accurately for all the codes allowing a detailed investigation of the
differences that develop due to the various gas physics
implementations employed. As expected, the mesh-based
codes \art\ and \arepo\ form extended entropy cores in
the gas with rising central gas temperatures. Those codes employing
traditional SPH schemes show falling entropy profiles all the way into
the very centre with correspondingly rising density profiles and
central temperature inversions. We show that methods with
modern SPH schemes that allow entropy mixing span the
range between these two extremes and the latest SPH variants produce gas entropy profiles that are essentially indistinguishable
from those obtained with grid based methods.
\end{abstract}
\noindent
\section{Introduction} \label{sec:introduction}
Galaxy clusters are the largest virialized objects in the Universe and, as 
such, provide both an important testbed for our theories of cosmological
structure formation and a challenging laboratory within which to study the 
fundamental physical processes that drive galaxy formation and evolution. 
The overdense regions that contain clusters in the present-day Universe were
the first to collapse and virialize in the early Universe, and so our theories 
must predict their assembly history over a large fraction of the lifetime
of the Universe (see \citealt{Allen11} and \citealt{Kravtsov12} for a recent review).
 At the same time, galaxies in the cores of clusters have 
orbited within the often violently growing cluster potential over many 
dynamical times, while the broader cluster galaxy population is continually
replenished by the infall of galaxies from the field.

Computer simulations are now well established as a powerful and indispensable
tool in the interpretation of astronomical observations (see for instance \citealt{Borgani11}). Early
$N$-body simulations (\citealt{White76}; \citealt{Fall78}; \citealt{Aarseth79}), which included the gravitational effects of dark matter
only, were vital in interpreting the results of galaxy redshift surveys and
unveiling the large scale structure of the Universe, and subsequently in
resolving structure in the non-linear regime of dark matter halo formation. 
The focus of modern simulations has now shifted to modeling galaxy formation
in a cosmological context (see \citealt{Borgani11} for a recent review), incorporating the key physical processes that
drive galaxy formation -- such as the cooling of a collisional gaseous 
component (e.g. \citealt{Pearce00}; \citealt{Muanwong01}; \citealt{Dave02}; \citealt{Kay04}; \citealt{Nagai07}; \citealt{Wriesma09}), the birth of stars 
from cool overdense gas (e.g. \citealt{Springel03}; \citealt{Schaye08}), the growth of black
holes \citep{DiMatteo05}, and the injection of energy into the inter-stellar medium by supernovae (e.g. \citealt{Metzler94}; \citealt{Borgani04}; \citealt{Dave08}\citealt{dallavecchia12}) and powerful AGN outflows (e.g. \citealt{Thacker06a}; \citealt{Sijacki07}; \citealt{Puchwein08}; \citealt{Sijacki08}; \citealt{Booth09}). These processes span an enormous dynamic 
range, both spatial and temporal, from the sub-pc scales of black hole growth 
to the accretion of gas onto Mpc scales from the cosmic web. Galaxy clusters offer an ideal testbed for the study of 
these processes and their complex interplay, precisely because their enormous size 
encompasses the range of scales involved. For this 
reason, the study of galaxy formation and evolution in cluster environments 
occupies a fundamental position in observational and numerical astrophysics.

This raises the important question of how reliably cosmological galaxy
formation simulations recover the properties of galaxy clusters. In the now 
classic Santa Barbara Cluster Comparison Project, \cite{Frenk99} (SB99 from now on) compared the 
bulk dark matter and gas properties of a galaxy cluster formed in a non-radiative
cosmological hydrodynamical simulation run using twelve (then state-of-the-art) 
mesh- and particle-based (hereafter smoothed particle hydrodynamics, SPH) 
codes. They displayed visual representations of the cluster at $z=0$ and 
at $z=0.5$ when a major merger event was ongoing, and 
compared several observable quantities such as enclosed mass, gas temperature 
and X-ray emission. The large scatter in essentially all quantities between 
the twelve models is evident from the plots. The origin of these discrepancies 
was partly the poor timing agreement between the methods due to an ambiguity 
in the start redshift as well as large differences in the effective numerical 
resolution that arose because the simulation challenged the computing resources 
available at the time. The key discrepancy has, however, stood the test of times: 
the divergence between mesh-based and SPH codes in  
the radial entropy profile of the gas, defined in SB99 as
\begin{equation}
        \label{eq:entropy}
        S(R) =\log \left[T_{\rm gas}(R)/\rho_{\rm gas}(R)^{2/3}\right]
\end{equation} 
where $R$ is the spherical radius with respect to the cluster centre of 
mass; $T_{\rm gas}$ is the gas temperature; and $\rho_{\rm gas}$ is the 
gas density. Fig.18 of SB99 showed some tentative indication that the entropy 
profiles of the grid based codes (principally those of Bryan and Cen) displayed 
a constant entropy core whereas the entropy profiles of the sparticle 
based SPH codes continued to fall all the way into the centre. 

These results have been confirmed subsequently by several studies 
\citep{Voit2005, OShea05, Dolag2005, 2008MNRAS.387..427W, 2009MNRAS.395..180M}. For example,
\citet{2008MNRAS.387..427W} and \citet{2009MNRAS.395..180M} suggested that the 
discrepancy owes also to the artificial surface tension and the associated lack
of multiphase fluid mixing in classic SPH, while similar conclusions have been
reached by \citet{2012MNRAS.424.2999S} when comparing the moving mesh code
\arepo\ of \citet{Springel10} with classic SPH, using P-GADGET3 with the entropy conserving SPH
version of \cite{Springel2002}.
Interestingly, in their recent study, \citet{Power14} have shown 
that SPHS \citep{2012MNRAS.422.3037R}, an extension of SPH that improves 
the treatment of mixing by means of entropy dissipation, produces constant entropy 
cores in non-radiative galaxy cluster simulations that are consistent with the 
results of the adaptive mesh refinement (AMR) code \ramses\
\citep{2002A&A...385..337T}. Both \citet{2008MNRAS.387..427W} and 
\citet{2009ApJ...707...40M} report entropy cores in runs that incorporate
sub-grid models for turbulence. These results suggest that modern SPH
codes can overcome the problems first reported in \citet{Frenk99} and 
subsequently in \citet{Agertz07}. It is worth noting that it is
not obvious that mesh-based codes necessarily recover the correct form for 
the entropy profile. For
example, \citet{Springel10} reports  
significant variation in the entropy profile for the same AMR 
code (\enzo) that is particularly sensitive to choice of refinement
criteria. More generally, differences are apparent when comparing
AMR results to that of the moving mesh code \arepo\;
\citet{Springel10} report an entropy core that is significantly 
lower than that found in AMR codes (e.g. compare Figure 45 of 
\citealt{Springel10} with Figure 18 of 
\citealt{Frenk99} or Figure 5 of 
\citealt{Voit2005}).

In this work - emerging out of the `nIFTy cosmology' workshop\footnote{\url{http://popia.ft.uam.es/nIFTyCosmology}} - we revisit the idea of the Santa Barbara Cluster Comparison Project fifteen 
years later. We take four
modern cosmological simulation codes (with one of them taken in nine different flavors, for
a total of twelve different codes) and study
the formation and evolution of a large galaxy cluster (with final mass 
$M_{\rm 200}^{\rm crit} \simeq 1.1 \times 10^{15}$M$_\odot$). First we perform blind dark 
matter only simulations with the favored parameter sets of each group. The 
results from these show the typical scatter that is expected for 
currently published works in this area. We then use a common parameter set to 
align our gravitational framework before continuing to study non-radiative gas simulations.
This allows us to focus solely on the 
differences between the models that arise due to the different hydrodynamical 
implementations. This also permits us to cleanly study the formation (or not) of a 
gas entropy core.

The rest of this paper is organized as follows: in Section~\ref{sec:codes} we 
briefly describe the twelve methods used in this study and supply tables 
listing parameter choices. In Section~\ref{sec:data} we describe how 
our initial conditions were generated and chosen. The main results are 
presented in Section~\ref{sec:DMcomparison}, which discusses the dark matter 
only simulations, and in Section~\ref{sec:NRcomparison}, which presents the results 
from the non-radiative simulations. We discuss convergence and scatter among the different codes
in Section~\ref{sec:scatter}. We summarize out results in 
Section~\ref{sec:conclusion}. We present additional supporting material the Appendix.

\section{The Codes} \label{sec:codes}

\begin{table}
  \caption{List of all the simulation codes participating in the nIFTy cluster comparison project.}
\label{tab:codes}
\begin{center}
\begin{tabular}{ll}
\hline
Code name				& Reference\\
\hline
CART				& \cite{Rudd08}\\
\arepo					& \cite{Springel10}\\
\hydra					& \cite{Couchman95}\\
\\
\gadget:				& \cite{Springel05}\\
G2-Anarchy	                & Dalla Vecchia et al. {\rm in prep.} \\
G3-X				& \cite{Beck15}\\
G3-SPHS				& \cite{2012MNRAS.422.3037R}\\
G3-Magneticum		& \cite{Magn14}\\
G3-PESPH			& Huang et al. {\rm in prep.}\\
G3-MUSIC				&\cite{Sembolini2013} \\
G3-OWLS				& \cite{Schaye10}\\
G2-X				& \cite{Pike14}\\
\hline
\end{tabular}s
\end{center}
\end{table}


The numerical codes used for this project can be divided into two main groups: mesh-based and SPH codes.
The mesh based codes used in this work are \art\ \citep{Kravtsov97} and \arepo\ \citep{Springel10}: the first one uses Eulerian hydrodynamics as the second one uses a moving unstructured mesh and can be considered almost Lagrangian. SPH codes use Lagrangian hydrodynamics: we used \hydra\ \citep{Couchman95} and nine different versions of \gadget\ \citep{Springel05}. Among  SPH codes we can distinguish {\rm classic} and {\rm modern} SPH, defining the latter as codes that adopted an improved treatment of discontinuities.
The codes employ different techniques to solve the evolution equations for a
two-component fluid of dark matter and non-radiative gas
coupled by gravity. To calculate gravitational forces, \art\ uses Adaptive Mesh Refinement (AMR), \arepo\ and \gadget\ are based on TreePM (Tree+Particle-Mesh) methods and \hydra\ uses 
AP$^3$M (Adaptive Particle-Particle, Particle-Mesh). Gas particles are treated in the following ways: by means of SPH in \gadget\ and \hydra\ , using a Voronoi mesh in Arepo, and using Eulerian AMR in ART.

The following short summaries detail the specifics of each simulation code contributing to this comparison (the reference author for each code is the person who
run the simulation). We focus on key differences and novel aspects. Generalized descriptions of the individual codes can be found in their respective methods papers. Table \ref{tab:codes} provides the standard reference for each code; Table \ref{tab:globalparams} summarizes the key numerical parameters used for each run. Sec. \ref{sec:modernSPH} describes the main improvements made by modern SPH codes.

\subsection{Mesh-based codes}
\noindent 
\paragraph*{\bf CART } (Nagai, Nelson \& Lau) 

\art\ (Adaptive Refinement Tree, ART) is an N-body plus hydrodynamics
code with adaptive mesh refinement. The code solves the inviscid fluid
dynamical equations using a second order accurate Godunov method with
piecewise-linear reconstructed boundary states and the exact Riemann
solver of \cite{Colella1985}. The current version of the code used for this work is CART (Chicago-ART), which
it is parallelized for distributed machines using MPI and features a
flexible time-stepping hierarchy.

\noindent 
\paragraph*{\bf \arepo\ } (Puchwein)

Arepo uses a Godunov scheme on an unstructured moving Voronoi mesh. The mesh cells move (roughly) with the fluid. The main differences to Eulerian AMR codes consist in that AREPO is almost Lagrangian and it is Galilean invariant by construction; furthermore, AREPO has automatic refinement for hydrodynamics and gravity and uses a Tree-PM gravity solver. The main difference to SPH codes is that the hydrodynamic equations are solved with a finite-volume Godunov scheme. In this work, we always use the total energy as a conserved quantity in the Godunov scheme rather than the entropy-energy formalism described in \cite{Springel10}.
\subsection{SPH codes}

\noindent 
\paragraph*{\bf \gadgetanarchy} (Dalla Vecchia) 

Gadget-Anarchy (G2-Anarchy) is an implementation of \gadget 2 employing the pressure-entropy SPH formulation derived by \cite{Hopkins13}. The artificial viscosity switch has been been implemented following \cite{Cullen10}, whose algorithm allows precise detection of shocks and avoid excessive viscosity in pure shear flows.
G2-Anarchy uses a purely numerical switch for entropy diffusion similar to the one of \cite{Price2008}, but without requiring any diffusion limiter.
The kernel adopted is the $C^2$ function of \cite{Wendland95} with 100 neighbors, with the purpose of avoiding particle pairing (as suggested by \citealt{Dehnen2012}).
The time stepping adopted is described in \cite{Durier12}.

\noindent 
\paragraph*{\bf \gadgetx} (Murante) 

This project employs two versions of GADGET3-X (G3-X). The default code, GADGET3-X-Std (G3X-Std) is a standard version of \gadget 3 with the cubic spline kernel, 64 neighbours, low-order viscosity and a shear flow limiter \citep{balsara95} and no conduction.
The improved version GADGET3-X-ArtCond (G3X-Art; \citealt{Beck15}) is an advanced version of \gadget 3 with the Wendland kernel functions \citep{Dehnen2012} with 200/295 neighbours, artificial conductivity to promote fluid mixing following \cite{Price2008} and \cite{tricco13}, but with an additional limiter for gravitationally induced pressure gradients, a time-step limiting particle wake-up scheme \cite{pakmor12} and a high-order scheme for gradient computation \citep{Price2012}, shock detection and artificial viscosity similar to \cite{Cullen10} and \cite{hu14}.
A companion paper \citep{Beck15} presents an improved hydrodynamical scheme and its performance in a wide range of test problems.
\noindent 
\paragraph*{\bf \gadgetsphs} (Power) 

\gadgetsphs\ (G3-SPHS) was developed to overcome the inability of classic SPH to resolve instabilities. It has been implemented in the \gadget 3 code. The key differences with respect to standard \gadget 3 are in the choice of kernel and in dissipation. 
Rather than the cubic spline kernel, G3-SPHS uses as an alternative either the HOCT kernel with 442 neighbors or the Wendland C4 kernel with 200 neighbors. 
A higher order dissipation switch detects, in advance, when particles are going to converge. If this happens, conservative dissipation is switched on for all advected fluid quantities. The dissipation is switched off again once particles are no longer converging.
This ensures that all fluid quantities are single-valued throughout the flow by construction.

\noindent 
\paragraph*{\bf \gadgetmagneticum} (Saro) 

Magneticum (G3-Magneticum) is based on the entropy-conserving formulation of SPH \citep{Springel2002}.  Higher order kernel based on the bias-corrected, sixth-order Wendland kernel \citep{Dehnen2012} with 295 neighbors.
Also included is a low viscosity scheme to track turbulence \citep{Dolag2005} and isotropic thermal conduction with $1/20^{th}$ Spitzer \citep{Dolag2004}.
\noindent 
\paragraph*{\bf \gadgetwindy} (February) 

\gadgetwindy\ (G3-PESPH) is an implementation of \gadget 3 with pressure-entropy
SPH \citep{Hopkins13} which features special galactic wind models. The SPH
kernel is an HOCTS (n=5) B-spline with 128 neighbors.  The time
dependent artificial viscosity is that of \cite{Morris1997}.
\noindent 
\paragraph*{\bf \gadgetmusic} (Yepes) 

The original MUSIC runs (G3-MUSIC) were done with the \gadget 3 code, based on the entropy-conserving formulation of SPH \citep{Springel2002}. \gadget 3 employs a spline kernel \citep{Monaghan85} and parametrize artificial viscosity following the model described by \cite{Monaghan97}. 
\noindent 
\paragraph*{\bf \gadgetowls} (McCarthy) 

The \gadgetowls\ (G3-OWLS) simulations were run using a version of the Lagrangian TreePM-SPH
code \gadget 3, which was
significantly modified to include new subgrid physics for radiative cooling, star formation, stellar feedback, black hole growth and AGN feedback, developed as part of the
OWLS/cosmo-OWLS projects \citep{Schaye10}. Standard entropy-conserving SPH \citep{lebrun14} was used with 48 neighbors.
\noindent 
\paragraph*{\bf \gadgettwox} (Kay) 

\gadgettwox\ (G2-X) is a modified version of the \gadget 2 code (SP05), using the
TreePM gravity solver and standard entropy-conserving SPH. It includes a number of sub-grid modules to model metal-dependent radiative cooling, star formation and feedback from supernovae and AGN.
\noindent 
\paragraph*{\bf \hydra} (Thacker) 

HYDRA-OMP (Hydra) \citep{Thacker06} is a parallel implementation of the earlier serial HYDRA code \citep{Couchman95}. Aside from the parallel nature of the code, HYDRA-OMP differs from the serial implementation by using the standard pair-wise artificial viscosity along with the Balsara modification for rotating flows. Otherwise, the SPH implementation is "classic" in nature, using 52 neighbors, and does not include more recently preferred kernels, terms for conduction or explicit shock tracking to modify the dissipation. It also uses a conservative time-stepping scheme that keeps all particles on the same synchronization.

\begin{table}
\caption{Key numerical parameters used for each run. Columns 2 and 3 list 
values for the Plummer-equivalent softening lengths for the DM particles 
in physical units; columns 4 and 5 the same but for the gas particles (where present); and
column 6 the number of FFT cells along each side of the box. We use the label 'Adp' when 
adaptive force resolution is used.}
\label{tab:globalparams}
\begin{center}
\begin{tabular}{llllll}
\hline
Code name & $\epsilon_{\rm DM}$ & $\epsilon^{\rm max}_{\rm DM}$ & $\epsilon_{\rm gas}$ & $\epsilon^{\rm max}_{\rm gas}$ & $N_{\rm FFT}$\\
\hline
CART					& Adp & Adp & Adp & Adp & 256\\
\arepo 					& 33.0 & 5.5 & Adp & Adp & 512\\
G2-Anarchy		        		& 20.0 & 5.0 & 20.0 & 5.0 & 512\\
G3-X						& 8.0 & 6.0 & 8.0 & 6.0 & 256\\
G3-SPHS					& 5.0 & 5.0 & Adp & Adp & 1024\\
G3-Magneticum			& 11.25 & 3.75 & 3.75 & 3.75 & 256\\
G3-PESPH				& 5.0 & 5.0 & 5.0 & 5.0 & 256\\
G3-MUSIC				& 8.0 & 6.0 & 8.0 & 6.0 & 512\\
G3-OWLS					& 9.77 & 5.0 & 9.77 & 5.0 & 1024\\
G2-X						& 24.0 & 6.0 & 24.0 & 6.0 & 256\\
\hydra 					& Adp & 5.0 & Adp & 5.0 & 512\\
\hline
\end{tabular}
\end{center}
\end{table}

\subsection{Progresses with modern SPH codes}
\label{sec:modernSPH}
Since the first development of SPH by \cite{gingold77} and \cite{lucy77} great advancements have been made to improve the accuracy and stability of SPH simulations.
In particular, much attention has been given to the treatment of discontinuities.
Artificial viscosity is necessary for a proper fluid sampling at shocks and to prevent particle interpenetration.
The first spatially constant low-order formulations of artificial viscosity applied viscosity not only in shocks, but also in shearing flows and un-shocked regions leading to an over diffusion of kinetic energy.
Modern formulations of artificial viscosity rely on proper shock detection methods and high-order gradient estimators to distinguish between shocked and un-shocked or shearing regions \citep{Morris1997,Cullen10,Price2012,hu14}.
Most importantly, they preserve kinetic energy to a much higher degree and promote simulations of turbulence or hydrodynamical instabilities.
Furthermore, SPH intrinsically fails to treat different gas phases and their mixing correctly, caused by the lack of diffusion terms and an always present spurious surface tension, as shown for instance by \cite{Agertz07}.

\begin{table*}
\caption{SPH schemata used for the comparison runs. We list the employed kernel functions and number of neighbours ($N_{\rm SPH}$) as well as the minimum smoothing length ($h_{\rm min}$) in terms of the gravitational softening length. Furthermore, we provides information about the artificial viscosity and conductivity switches.}
\label{tab:SPHparams}
\begin{center}
\begin{tabular}{lllllllll}
\hline
Code name & Hydrodyn. Kernel & $N_{\rm SPH}$ & $h_{\rm min}$ & Art. Visc. & Shearflow & Mixing & Limiter &\\
\hline
G2-Anarchy                        & Wendland C2 & $100\pm 3$ & 0 & Adaptive & LowOrder & Artificial & Yes \\
G3-XArt                 & Wendland C6 & $295\pm 10$ & 0.1& Adaptive & HighOrder & Artificial & Yes \\
G3-SPHS                            & Wendland C4 & $200\pm 0.5$ & 1.0 & Adaptive & LowOrder & Artificial & Yes \\
G3-Magneticum              & Wendland C6 & $295\pm0.5$ & 0.001& Adaptive & HighOrder & Physical & -\\
G3-PESPH                           & HOCTS B-spline & $128\pm 2$ & 0.1& Adaptive & LowOrder & - & - \\
G3-MUSIC                         & Cubic spline & $40\pm 3$ & 0.1& Constant & LowOrder & - & - \\
G3-XStd                            & Cubic spline & $64\pm 1$ & 0.1& Constant & LowOrder & - & - \\
G3-OWLS                            & Cubic spline & $48\pm 1$ & 0.01& Constant & LowOrder & - & - \\
G2-X                             & Cubic spline & $50\pm 3$ & 1& Constant & LowOrder & - & - \\
\hydra                                  & Cubic spline & $53\pm 1$ 5 & 0.5 & Constant & LowOrder & - & -\\
\hline
\end{tabular}
\label{tab:codes_imp}
\end{center}
\end{table*}

\cite{2009arXiv0906.0774R} showed that the mixing problem in SPH owes to two problems: the force inaccuracy and the lack of entropy mixing.
Artificial conduction \citep{Price2008} or pressure-entropy \citep{2001MNRAS.323..743R,saitoh13,Hopkins13} formulations have been developed to overcome these issues. 
They provide for the transport of heat between particles or change the basic physical variables.
However, in the presence of gravitationally induced pressure and temperature gradients, artificial conduction schemes might lead to unwanted transport of heat, making the use of numerical limiters as well as correction terms are highly desirable.
\cite{2012MNRAS.tmp.2941R} showed that pressure-entropy SPH fails for strong shocks and/or if the density gradient is large. This was significantly improved by \cite{Hopkins13} who derived a conservative pressure-entropy SPH for the first time. However, the problem with large density gradients remained. \cite{2012MNRAS.tmp.2941R} propose instead to use higher order switching, similarly to \cite{Cullen10}, but applied to all advected fluid quantities. This solved the mixing problem in SPH also for high density contrast and opened the door to "multimass" SPH for the first time.

Lastly, the commonly employed cubic spline kernel function can easily become unstable, which leads to the pairing instability, incorrect gradient estimators and a poor fluid sampling.
Therefore, a change of kernel function is highly recommended, and Wendland kernels \citep{Dehnen2012} are now commonly used to retain fluid sampling.
Table \ref{tab:SPHparams} provides an overview of the different SPH codes participating in our cluster comparison project and lists their numerical details.

\section{The Simulation} \label{sec:data}

The cluster we have adopted for this project was drawn from the MUSIC-2 sample
(\citealt{Sembolini2013}; \citealt{Sembolini2014}; \citealt{Biffi2014})
which consists of a mass limited sample \footnote{specifically, it is cluster 19 of MUSIC-2 dataset;
all the initial conditions of MUSIC clusters are available at \texttt{http://music.ft.uam.es}} of re-simulated halos selected
from the MultiDark cosmological simulation \citep{Prada12}.  This simulation is
dark-matter only and contains 2048$^3$ (almost 9 billions) particles in
a (1$h^{-1}$Gpc)$^3$ cube. It was performed in 2010 using ART
\citep{Kravtsov97} at the NASA Ames Research centre. All the
data from this simulation are accessible online through the {\itshape MultiDark
Database}.\footnote{www.MultiDark.org}  The run was done using the
best-fitting cosmological parameters to WMPA7+BAO+SNI ($\Omega_{\rm M} =
0.27$, $\Omega_{\rm b} = 0.0469$, $\Omega_{\Lambda}= 0.73$, $\sigma_8 =
0.82$, $n = 0.95$, $h = 0.7$, \citealt{Komatsu11}). This is the reference cosmological
model used in the rest of the paper.

The MUSIC-2 cluster catalogue was originally constructed by
selecting all the objects in the simulation box which are more massive
than 10$^{15}\;h^{-1}$M$_\odot$ at redshift $z=0$.  In total, 282
objects were found above this mass limit.  A zooming technique
described in \cite{Klypin01} was used to produce the
initial conditions for the re-simulations.  All particles within a
sphere with a radius of $6\;h^{-1}$Mpc around the centre of each selected object at
$z = 0$ were found in a low-resolution version ($256^3$ particles) of
the MultiDark volume. This set of particles was then mapped back to
the initial conditions to identify the Lagrangian region corresponding
to a $6\;h^{-1}$Mpc radius sphere centered on the cluster centre of
mass at $z = 0$. The initial conditions of the original simulations
were generated on a finer mesh of size $4096^3$. By doing so, the mass
resolution of the re-simulated objects was improved by a factor of 8
with respect to the original simulations. In the high resolution
region the mass resolution for the dark matter only simulations
corresponds to m$_{\rm DM}$=1.09$\times$10$^9\;h^{-1}$M$_\odot$. For the
runs with gas physics, m$_{\rm DM}$=9.01$\times$10$^8\;h^{-1}$M$_\odot$
and to m$_{\rm gas}$=1.9$\times$10$^8\;h^{-1}$M$_\odot$.

\section{Dark matter run comparison} \label{sec:DMcomparison}

\begin{figure*}
\centering
\begin{tabular}{ccc}
\includegraphics[width=50mm]{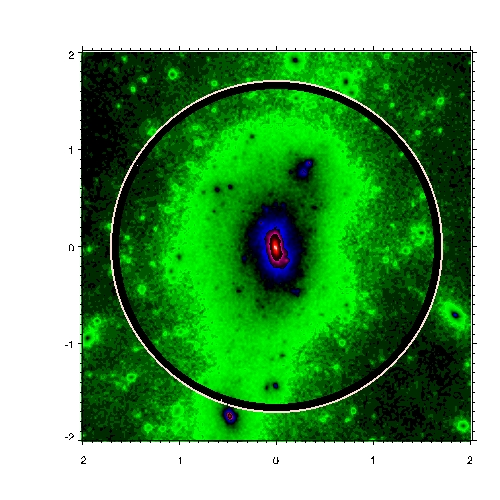}&
\includegraphics[width=50mm]{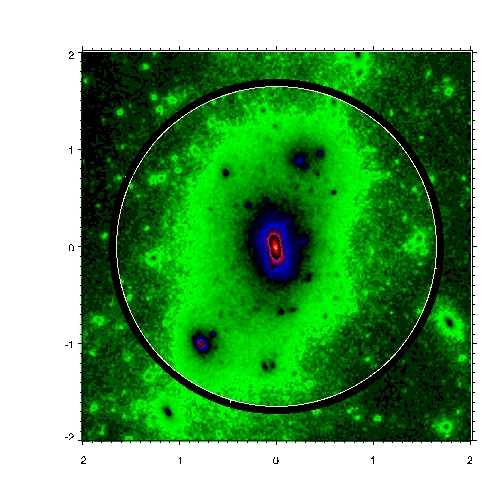}&
\includegraphics[width=50mm]{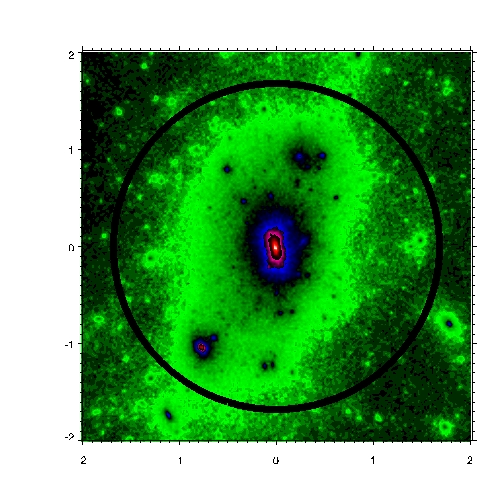}\\
CART & Arepo & G2-Anarchy\\
\includegraphics[width=50mm]{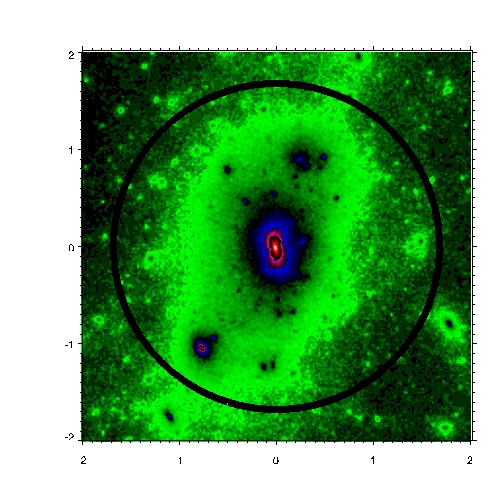}&
\includegraphics[width=50mm]{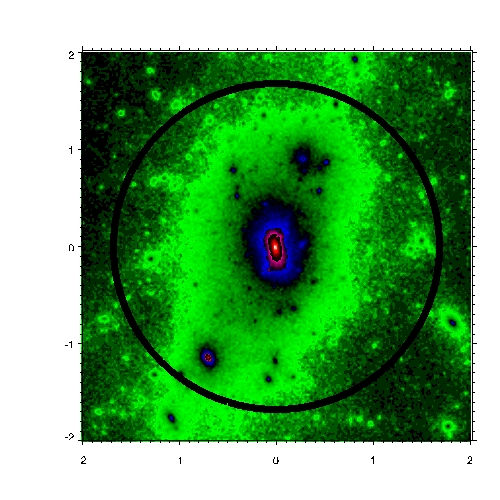}&
\includegraphics[width=50mm]{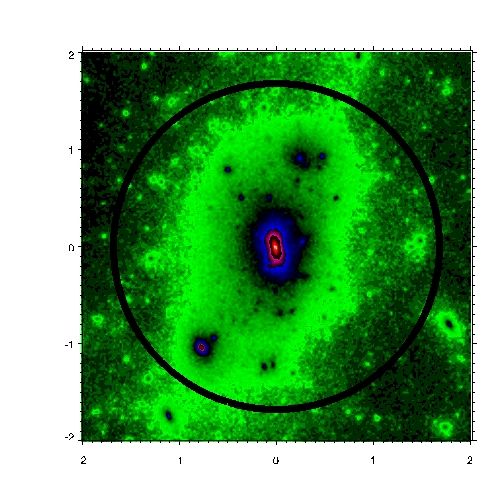}\\
G3-Mart & G3-SPHS & G3-Magneticum\\
\includegraphics[width=50mm]{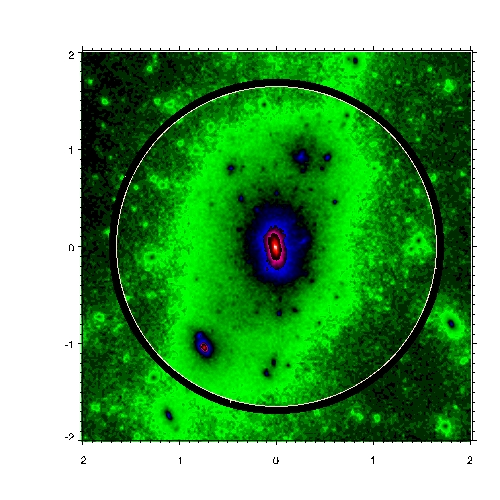}&
\includegraphics[width=50mm]{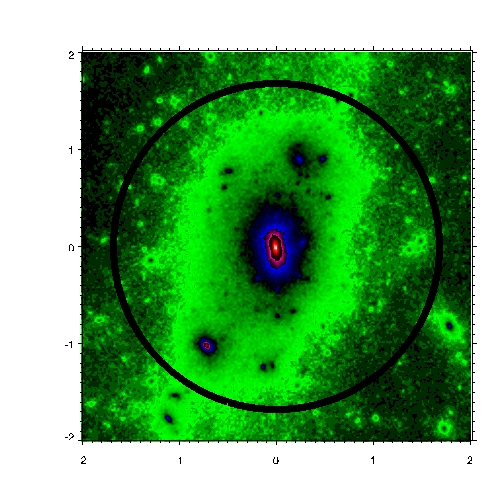}&
\includegraphics[width=50mm]{plots/new/dm/maps/jpg/GadgetX_2Mpc_z0_xy_circle_DM.jpg}\\
G3-PESPH & G3-MUSIC & G3-XStd\\
\includegraphics[width=50mm]{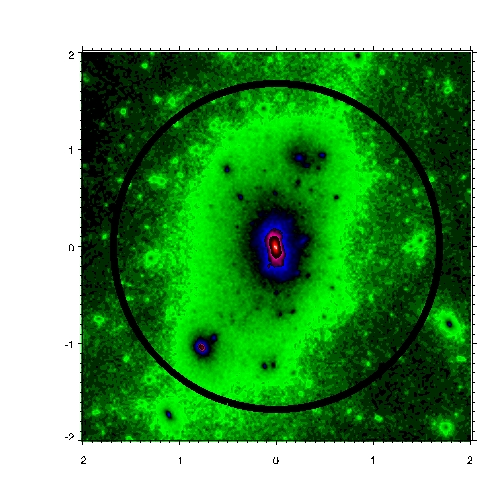}&
\includegraphics[width=50mm]{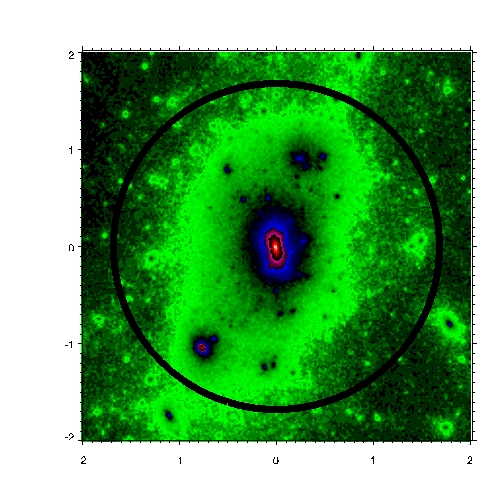}&
\includegraphics[width=50mm]{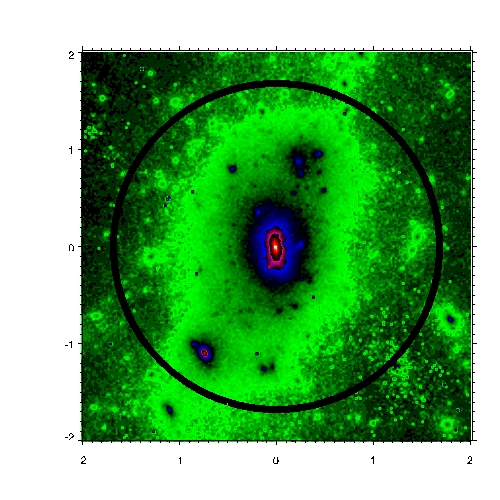}\\
G3-OWLS & G2-X & Hydra \\DM
\end{tabular}
\includegraphics[width=100mm]{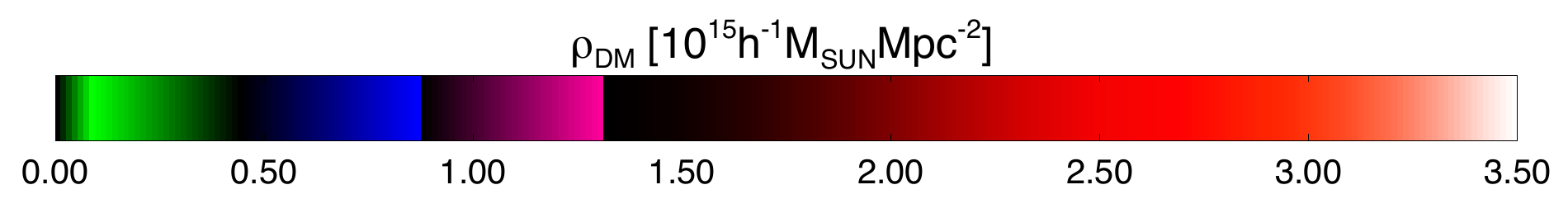} 
\caption{Projected density of the dark matter halo at $z=0$ for each simulation as indicated. The box is 2$h^{-1}$Mpc  on a side. The white circle indicates $M_{200}^{crit}$ for the halo, the black circle shows the same but for the G3-MUSIC simulation. }
\label{fig:DM_visual_z0}
\end{figure*}

We first completed a dark matter only simulation, performed using
the parameter values given in Table~\ref{tab:globalparams}. These were chosen independently by each
modeling group following their previous experience. A visual
comparison of the density field centered on the cluster at $z=0$ is
given in Figure~\ref{fig:DM_visual_z0}. While it is clear that all the
methods have followed the formation of the same object (with a significant improvement with
respect to Figure 1 of SB99) the precise
location of the major subhalo is not accurately recovered. For
several methods it has already crossed R$_{200}^{crit}$ (the radius enclosing
an overdensity of 200 relative to the critical density) while in
others it is still falling in. The variance across this figure
illustrates the typical range of outcomes from commonly used
cosmological codes. The major cause of the discrepancy (for the
\gadget\ based codes at least) is the size of the base level particle
mesh. Those implementations which employed a base mesh of 256$^3$ did not
sufficiently resolve the interface region between this low resolution mesh and the
higher resolution region placed over the cluster. Improving the
resolution of the base level mesh to 512$^3$ alleviates this problem and aligns
the dark matter component well. We demonstrate this by showing the
effect of changing the size of the particle mesh for the G3-MUSIC
code in the appendix (Figure \ref{fig:Non-cooling_visual_dm_z0}) together with a similar set of visual imagesÅ
taken from the following non-radiative simulation when the simulation
parameters influencing the accuracy of the dark matter are aligned
across the code. This demonstrates that, given a common set of
parameters, the dark matter framework underlying the simulation can be
made to look very similar (this is not surprising at least for the \gadget\ based codes and \arepo\, as they 
all use the same gravity solver).

\begin{figure*}
\includegraphics[width=.7\textwidth]{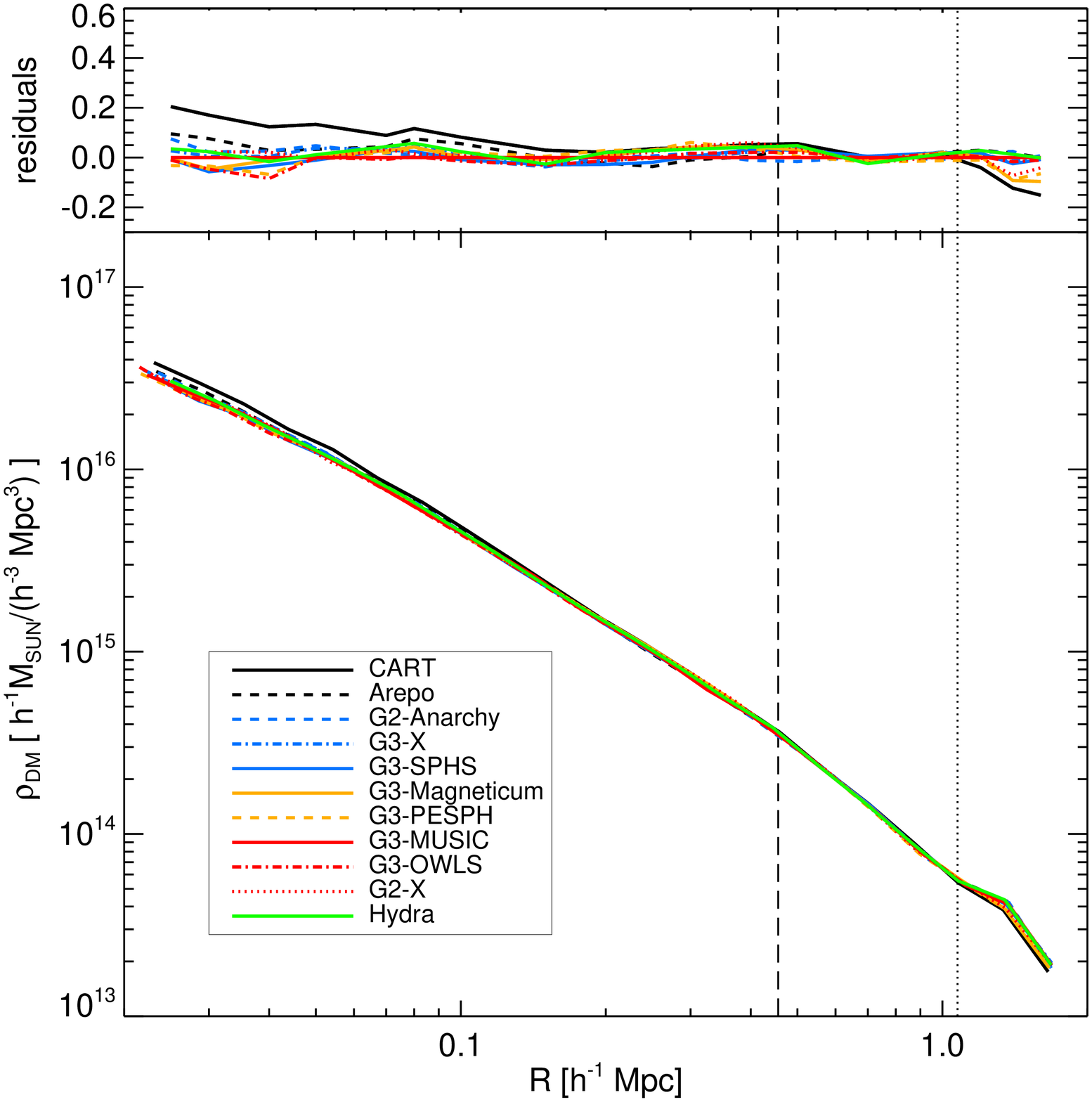}
\caption{Radial density profiles for the dark matter only simulations at $z=0$ (bottom panel) and difference between the radial density profiles of each dark matter only simulations at $z=0$ and the reference G3-MUSIC density profile (top panel). The vertical dashed line corresponds to $R_{2500}$ and vertical dotted line to $R_{500}$ of the reference G3-MUSIC values.}
\label{fig:DM_density_profile}
\end{figure*}

Figure~\ref{fig:DM_density_profile} displays the radially averaged projected dark
matter density profiles for the 12 different non-aligned DM-only runs, including also the residuals relative to
the density profile of the reference G3-MUSIC simulation. The
secondary peak marks the location of the major subhalo, at $R \sim$ 1$h^{-1}$Mpc,
 significantly closer to the centre in some simulations due to the size
of the top level particle mesh employed. All the simulations except
CART lie well within 10 per cent of each other at all radii with the \hydra\
simulation being indistinguishable from the \gadget\ runs. CART produces a
cluster that is slightly more centrally concentrated than for the particle
based approaches, especially within the inner 100$h^{-1}$kpc. 

\begin{figure*}
\includegraphics[width=.7\textwidth]{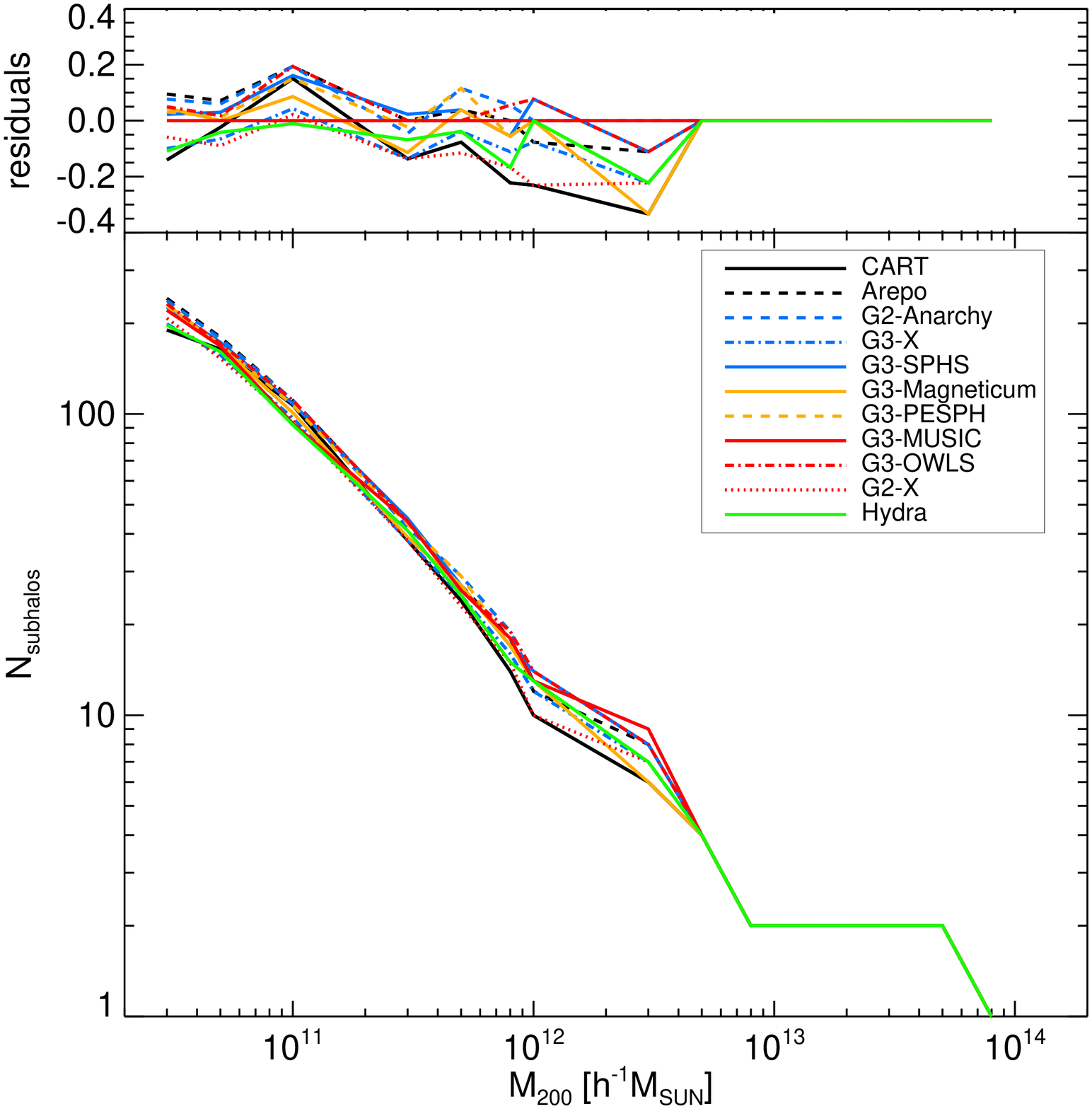}
\caption{Subhalo mass function for the dark matter only simulations at $z=0$ (bottom panel) and difference between the subhalo mass function  of each dark matter only simulations at $z=0$ and the reference G3-MUSIC subhalo count (top panel).}
\label{fig:DM_subhalo_massfunc_z0}
\end{figure*}

The subhalo mass function at $z=0$
(Figure~\ref{fig:DM_subhalo_massfunc_z0}) is recovered with a very close agreement (differences are always below 20 per cent at all masses) by all
codes. Subhalos here have been identified using AHF (\citealt{Gill04}; \citealt{Knebe09}; freely available from \texttt{http://popia.ft.uam.es/AHF}). The number of subhalos is essentially identical except for
tiny mass differences which are driven by the small divergences in
radial location that were identified above. These code to code variations lead to
differences in the mass associated with each subhalo as the threshold
that defines where a subhalo is separated from the background halo
varies. As expected, subhalos closer to the centre of the main halo
than the equivalent subhalo in one of the other models have a lower
recovered mass.

Comparison of the dark matter distribution and  of its radial
density profile at $z=1$ give results similar to those described
above. We conclude that the typical range of chosen parameters for
cosmological simulation codes introduces a variation of around 10 per cent
in the density profile of collapsed objects. This scatter can be reduced to less than 5 per cent by
aligning the numerical accuracy of our calculations. Although this is
not essential for many applications, we choose to do this in
the remainder of this paper so that, as we show in the appendix, the
underlying dark matter framework agrees closely, allowing
us to focus on differences resulting from the various
hydrodynamical schemes.

\section{Non-radiative run comparison} \label{sec:NRcomparison}
\begin{figure*}
\includegraphics[width=0.87\textwidth]{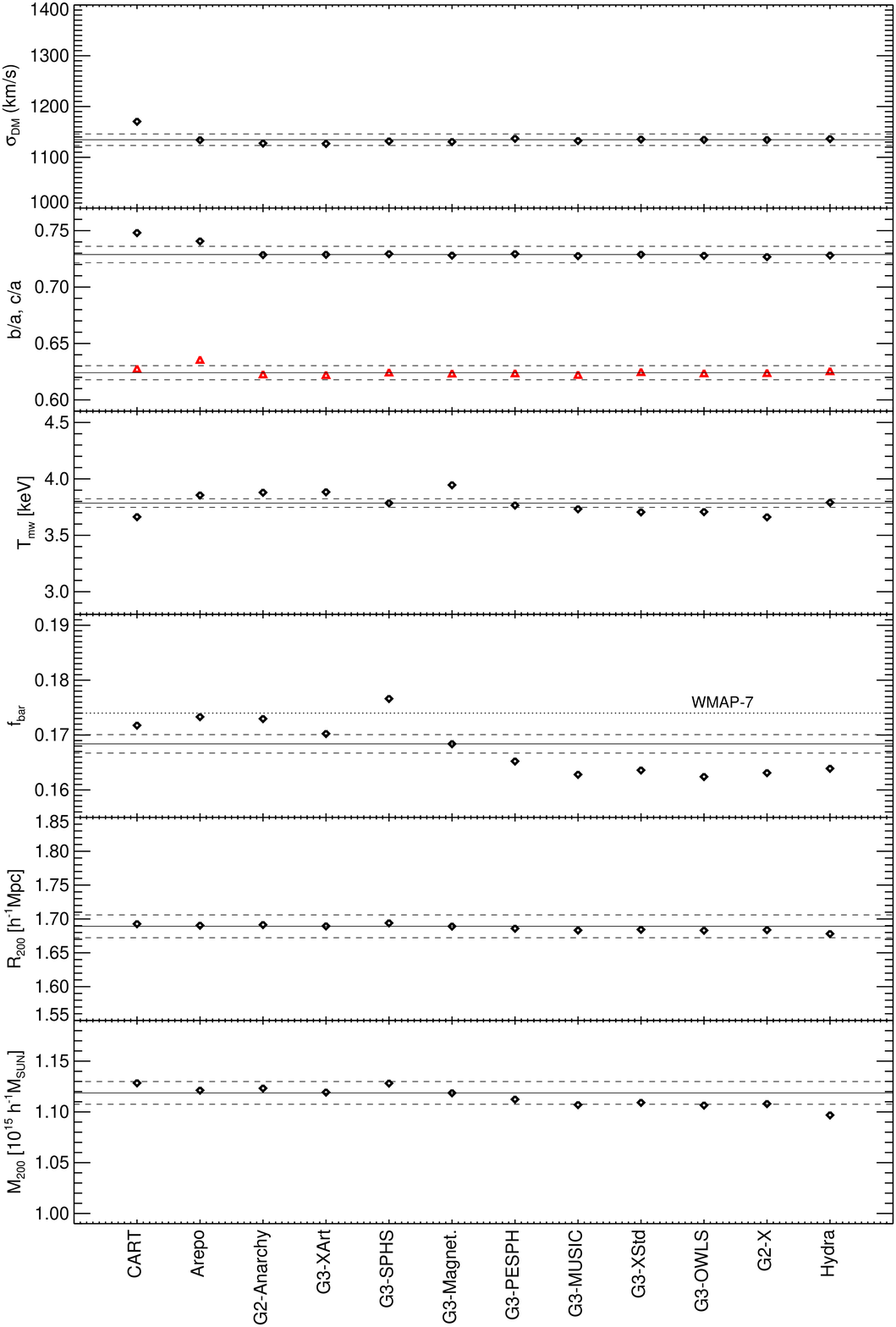}
\caption{Global properties of the cluster calculated by the different codes. All quantities are computed within $R_{200}^{crit}$. From top panel to bottom panel: $(1)$ the one-dimensional dispersion of the dark matter, $(2)$  the axial ratio ($b/a$ in black, $c/a$ in red), $(3)$ the mass-weighted temperature,  $(4)$ the gas fraction (dotted line represented the value of the cosmic ratio from WMAP7 \citep{Komatsu11}, $(5)$ the radius and $(6)$ the total cluster mass. The solid lines represent the median value for each one of the plotted quantities and the dashed lines the +/- 1 per cent scatter.}
\label{fig:global}
\end{figure*}

We now proceed to include a gas phase into our calculations. We repeat
the simulation of the same cluster as used in Section \ref{sec:DMcomparison} including gas
which however we do not allow to radiate energy. 
Figure \ref{fig:global} shows some of the global properties of the selected cluster as calculated by the different codes used in this work: radius, mass, mass-weighted temperature, gas fraction, dark matter velocity dispersion and axial ratios.
All these quantities have been calculated at an aperture radius corresponding to $R_{200}^{crit}$, the radius enclosing an overdensity of 200 relative to the critical density, defined as 
 \beq
\rho_c(z)=\frac{3H_0^2E(z)^2}{8\pi G}
\eeq
where {\itshape H$_0$} is the present value of the Hubble constant, G is the universal gravitational constant -- using the same definition we refer in the text to $R_{2500}^{crit}$ and $R_{500}^{crit}$ as the radii enclosing an overdensity of 2500 and 500 to $\rho_c$($z$).
It is interesting to note that all the codes were able to recover the same values for the different properties of the halo with a scatter smaller than 1 per cent for mass, radius, axial ratio and dark matter velocity dispersion and around 2 per cent for baryon fraction and gas temperature. The same properties were estimated with a scatter between 5 and 10 per cent in SB99, with differences of up to 30 per cent between the maximum and minimum values: in our comparison the same difference is always below 5 per cent (only for the gas fraction we register a disagreement of 8 per cent between the maximum and minimum values).

\begin{figure*}
\centering
\begin{tabular}{ccc}
\includegraphics[width=50mm]{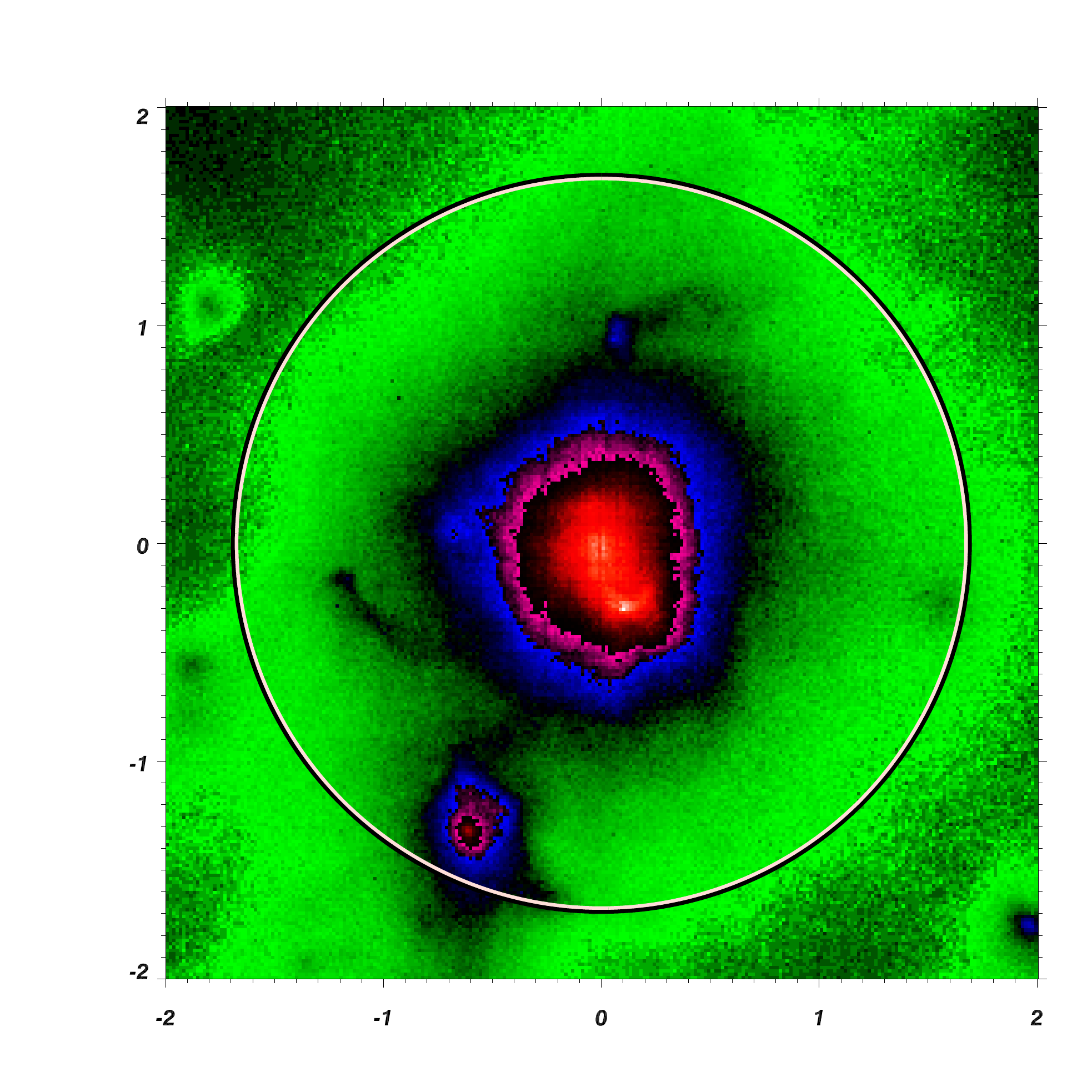}&
\includegraphics[width=50mm]{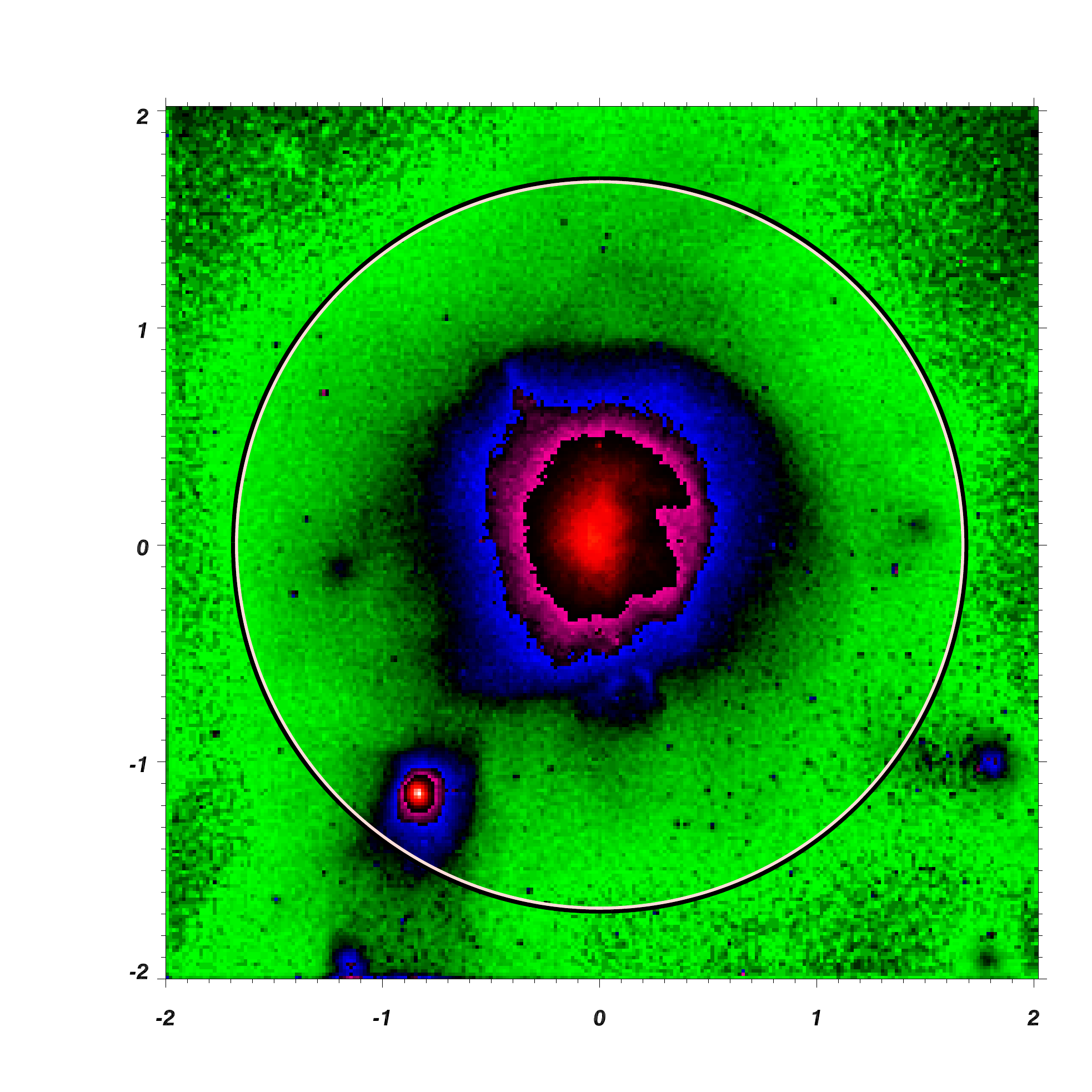}&
\includegraphics[width=50mm]{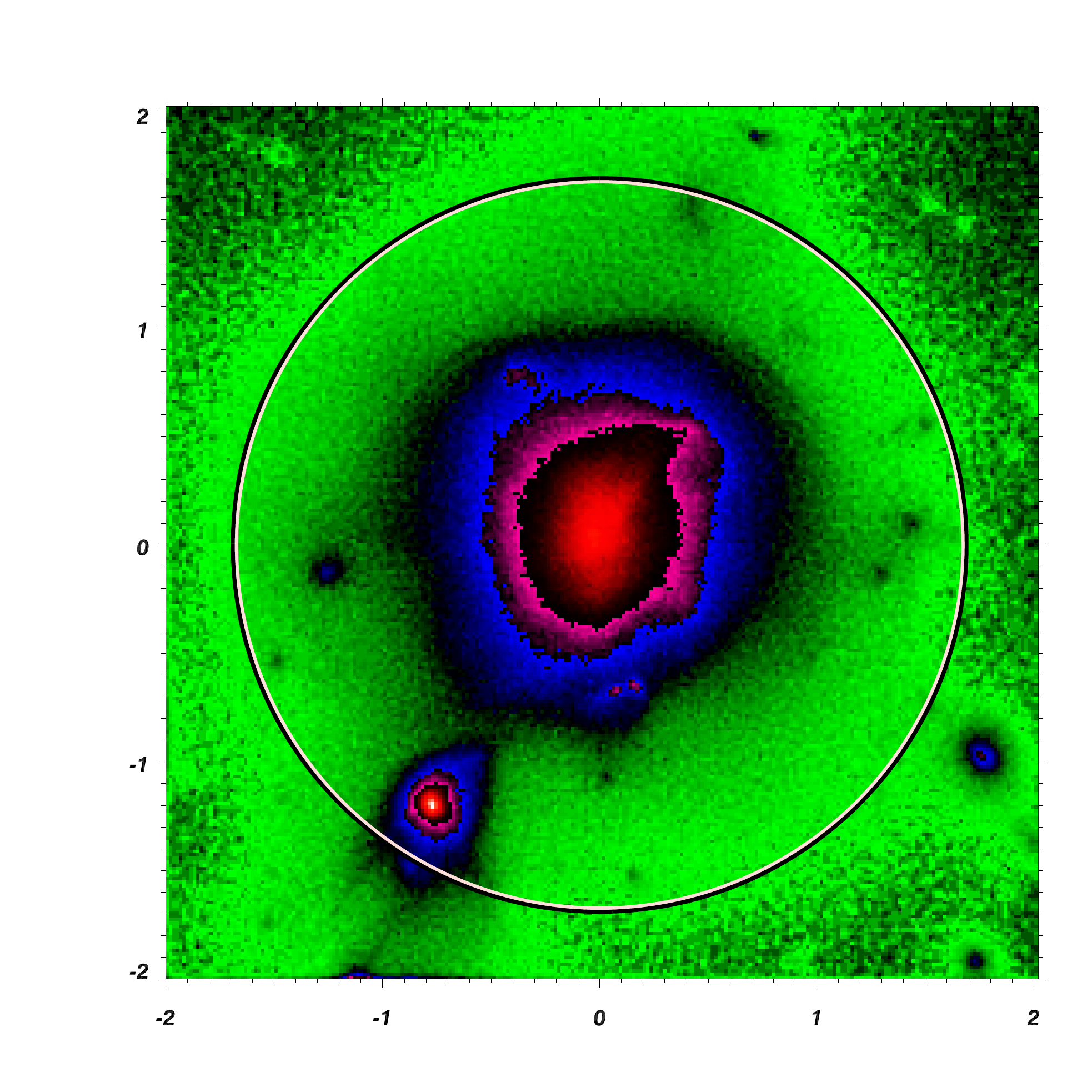}\\
ART & Arepo & G2-Anarchy\\
\includegraphics[width=50mm]{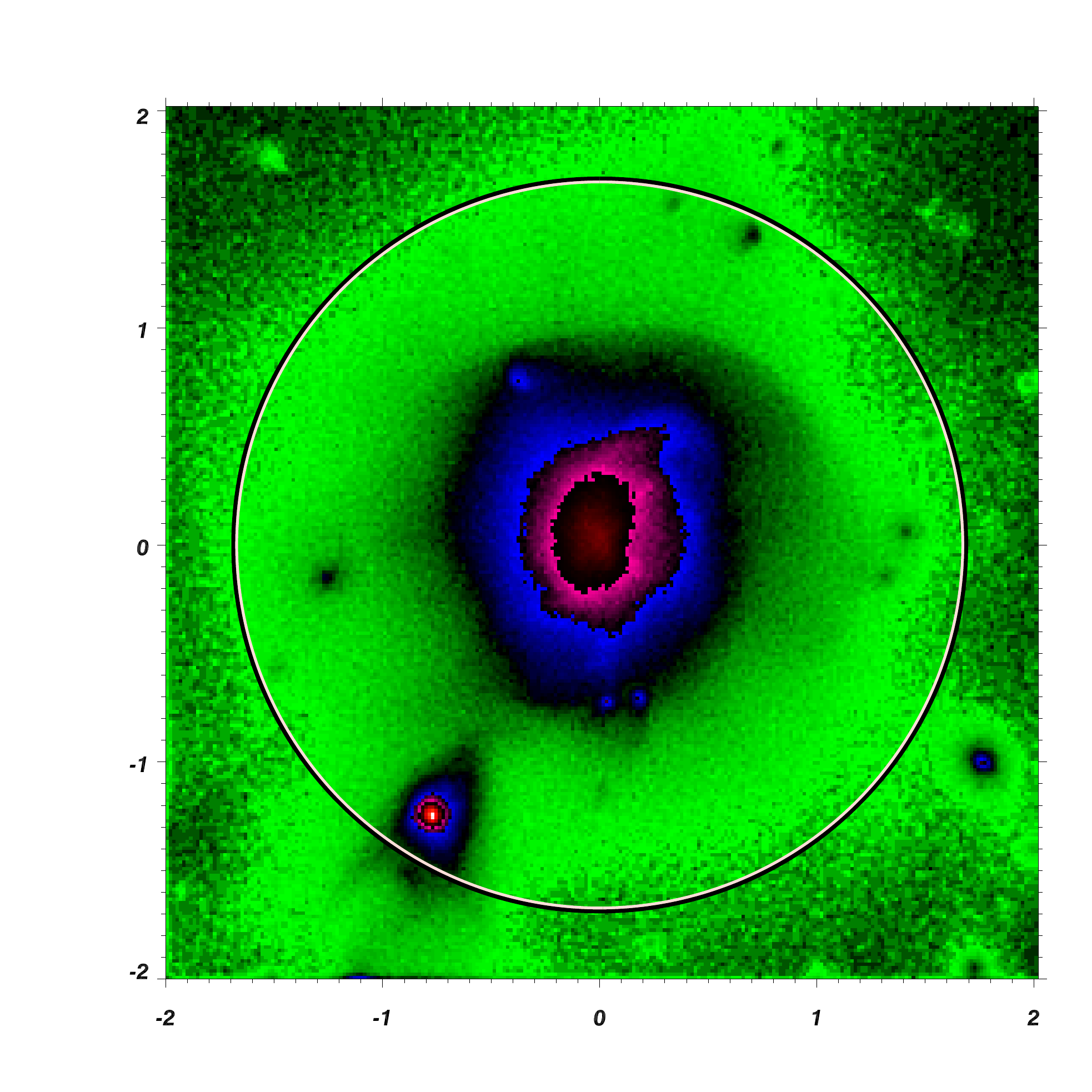}&
\includegraphics[width=50mm]{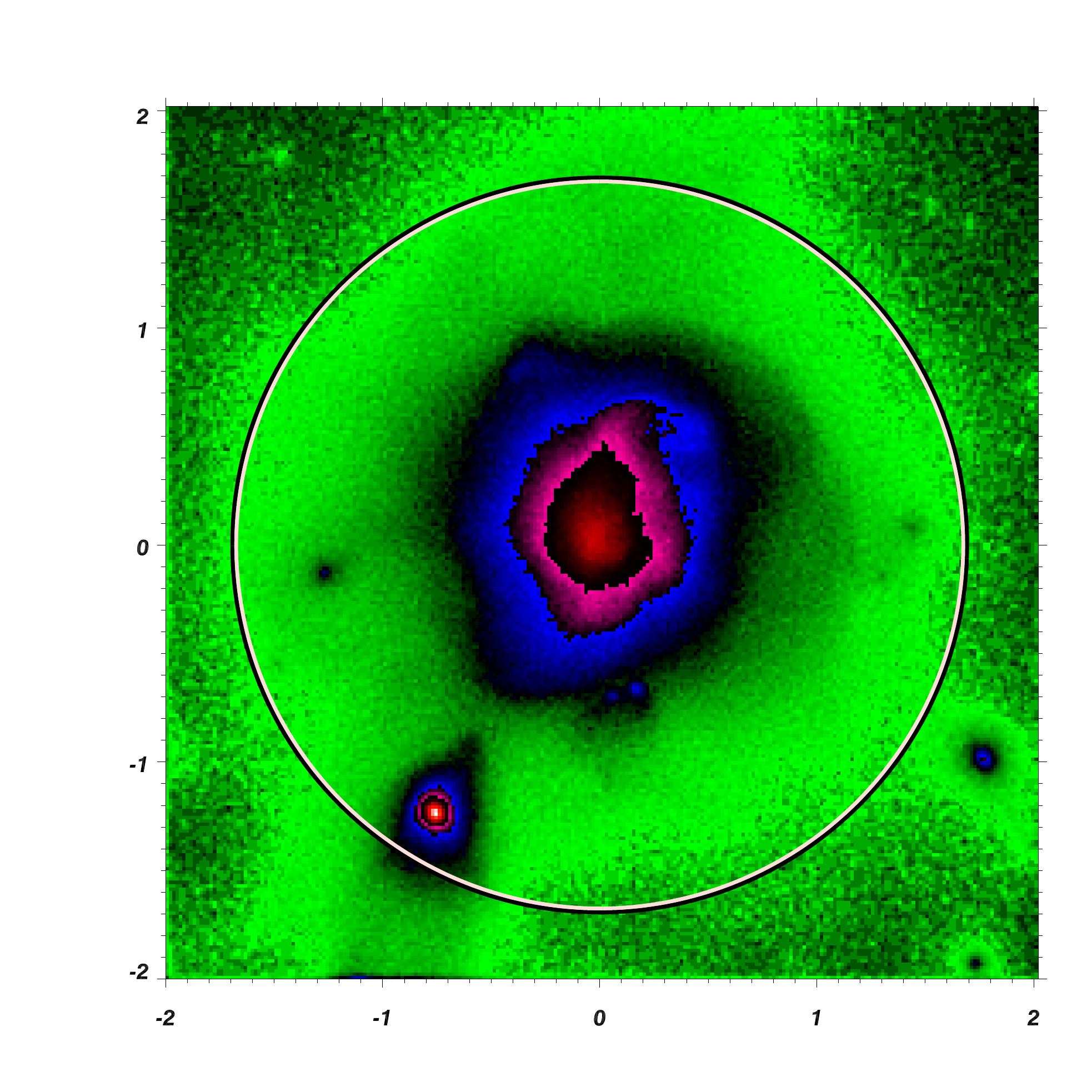}&
\includegraphics[width=50mm]{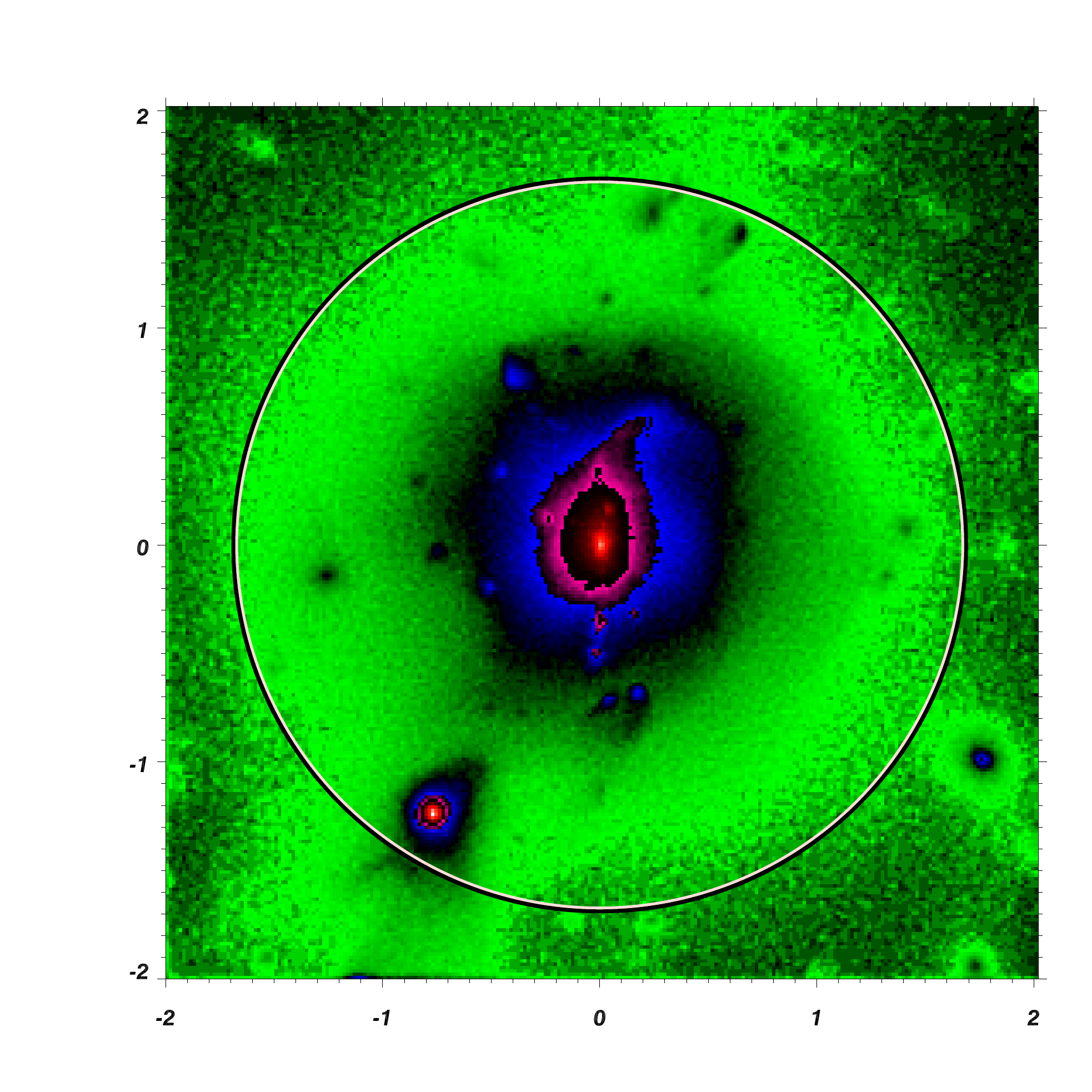}\\
G3-XArt & G3-SPHS & G3-Magneticum\\
\includegraphics[width=50mm]{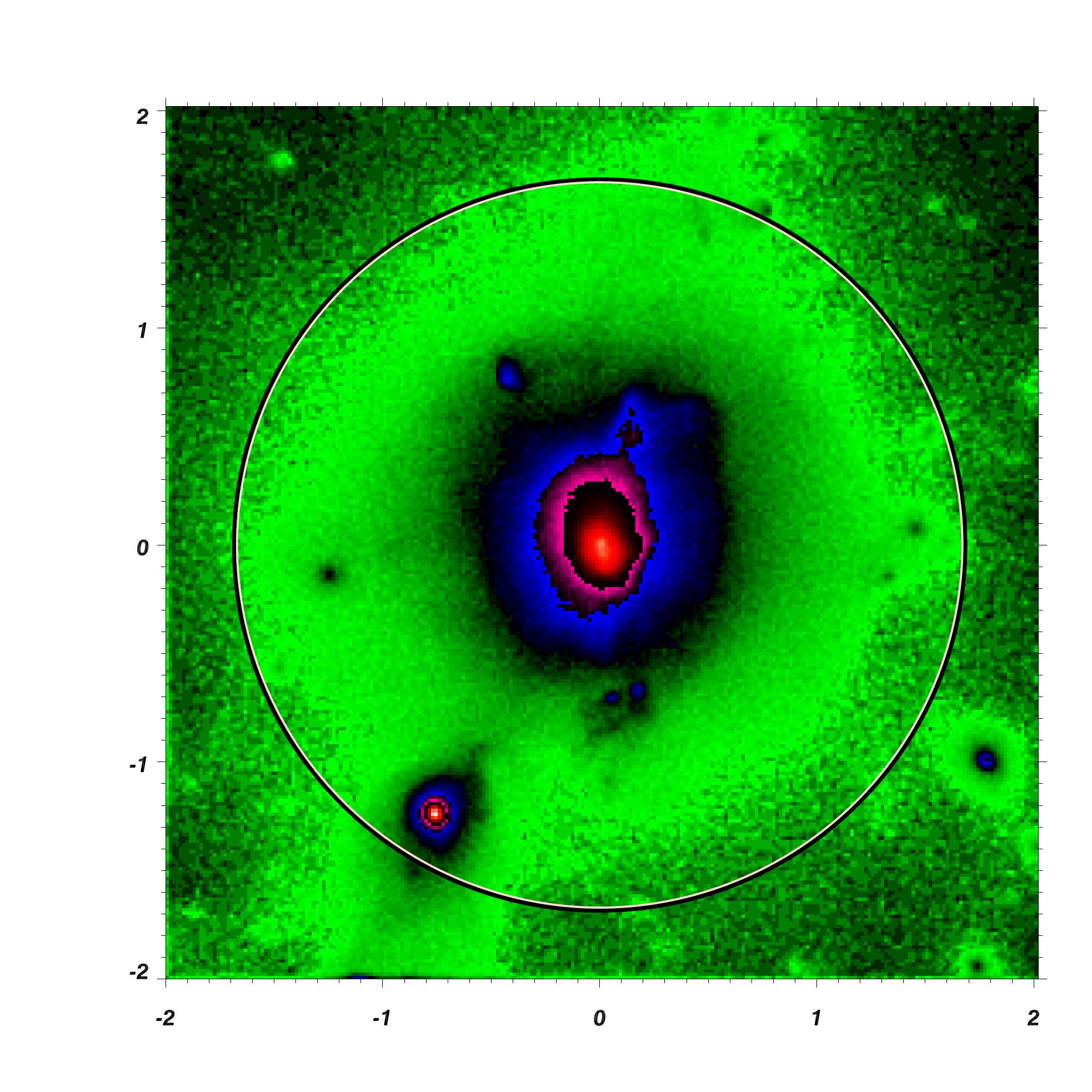}&
\includegraphics[width=50mm]{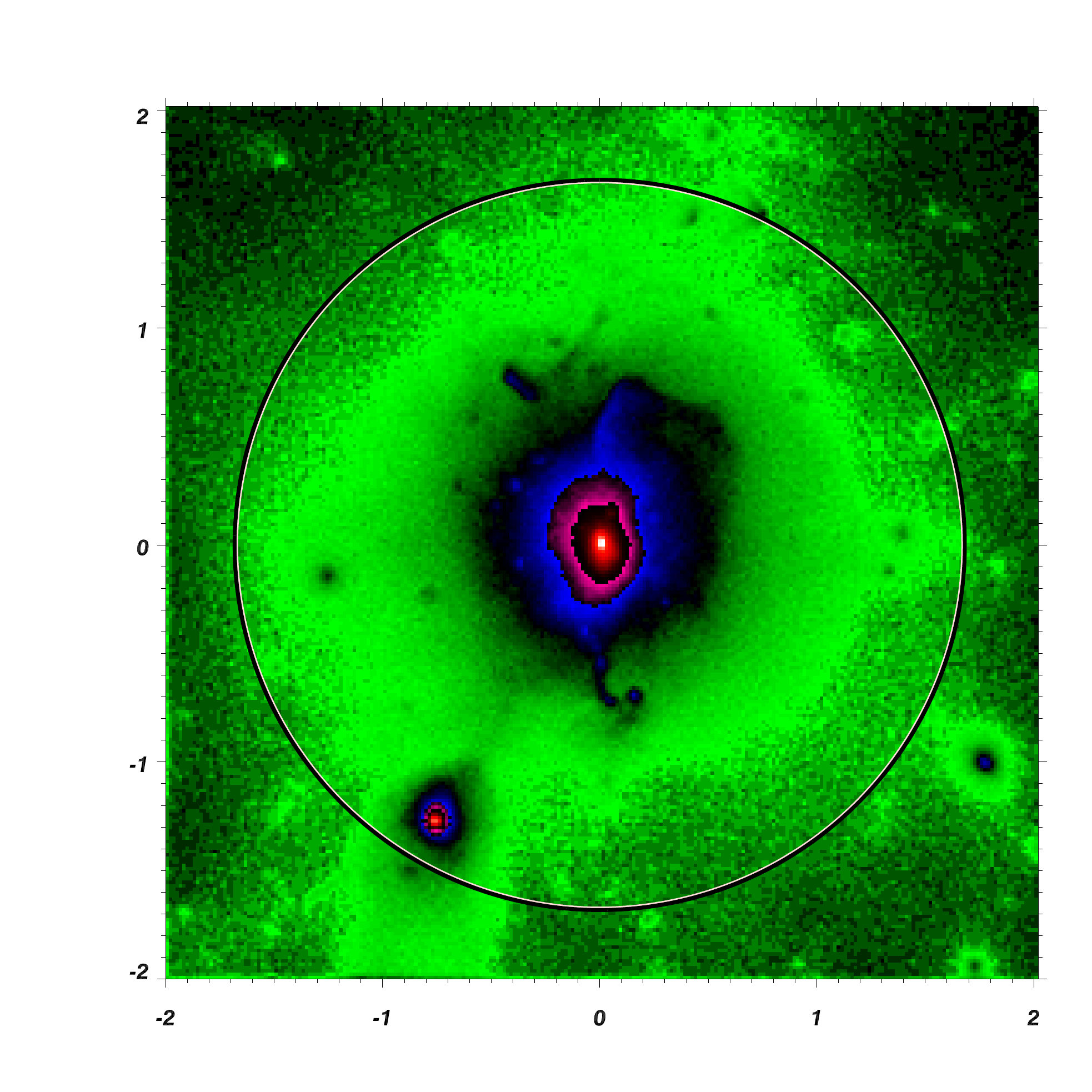}&
\includegraphics[width=50mm]{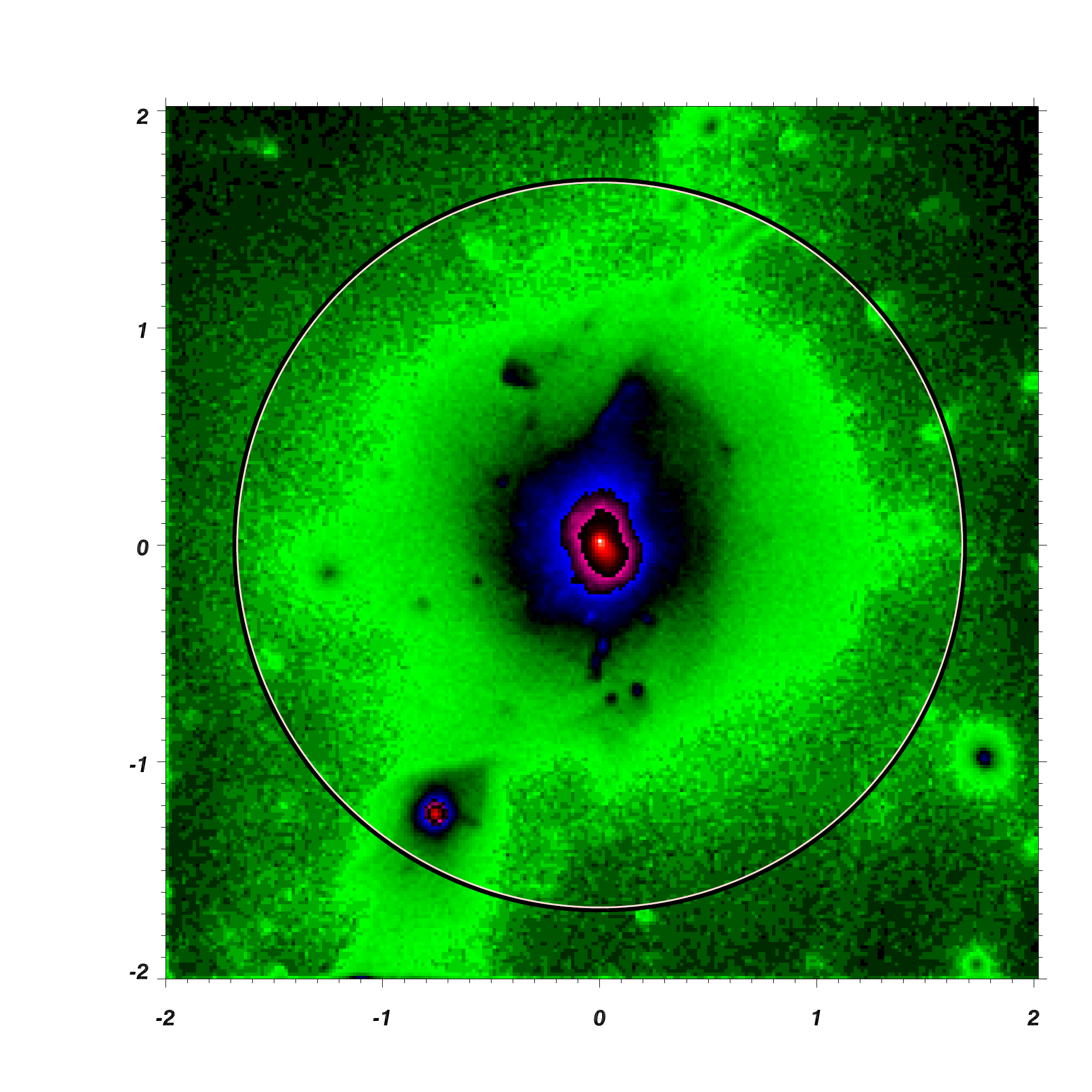}\\
G3-PESPH & G3-MUSIC & G3-XStd\\
\includegraphics[width=50mm]{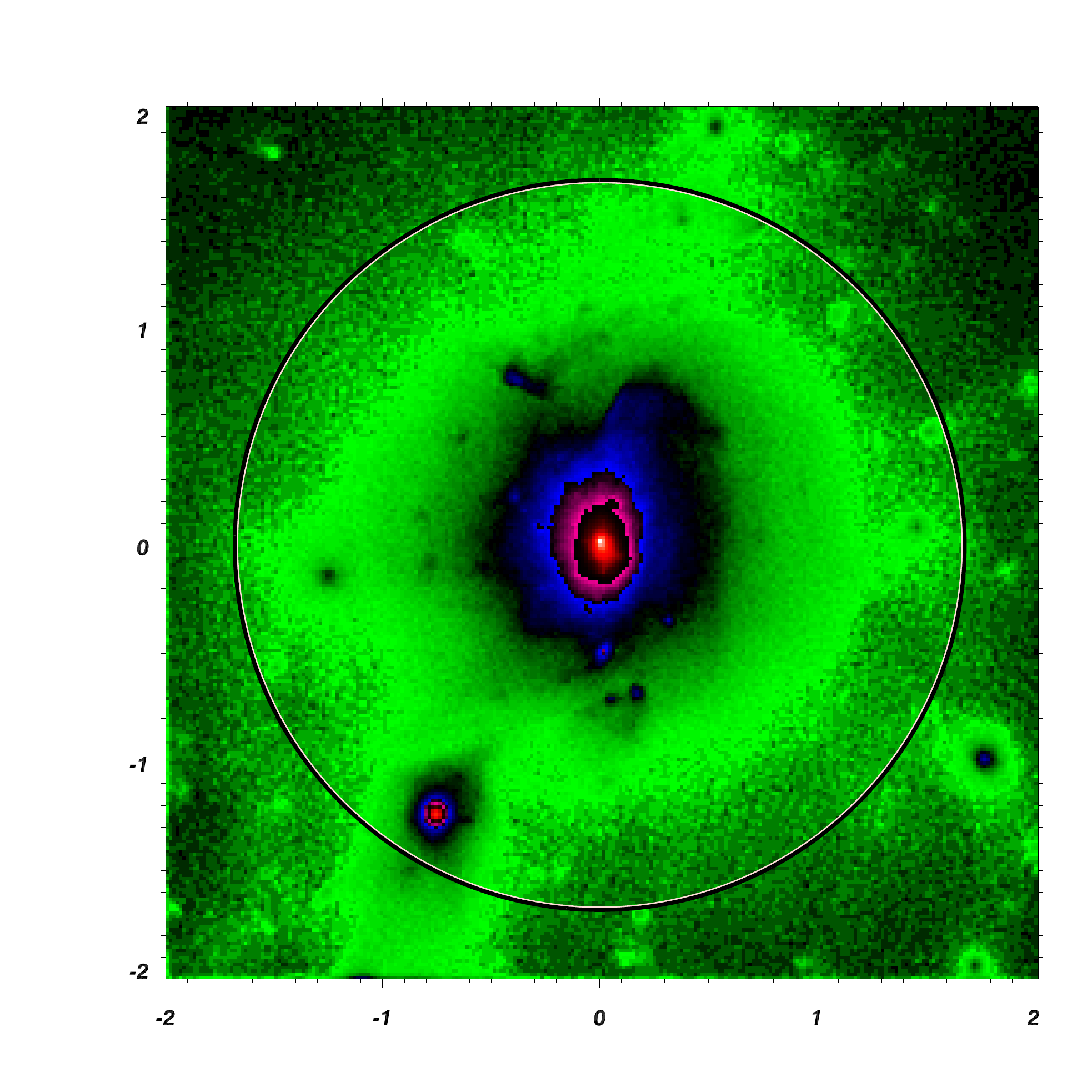}&
\includegraphics[width=50mm]{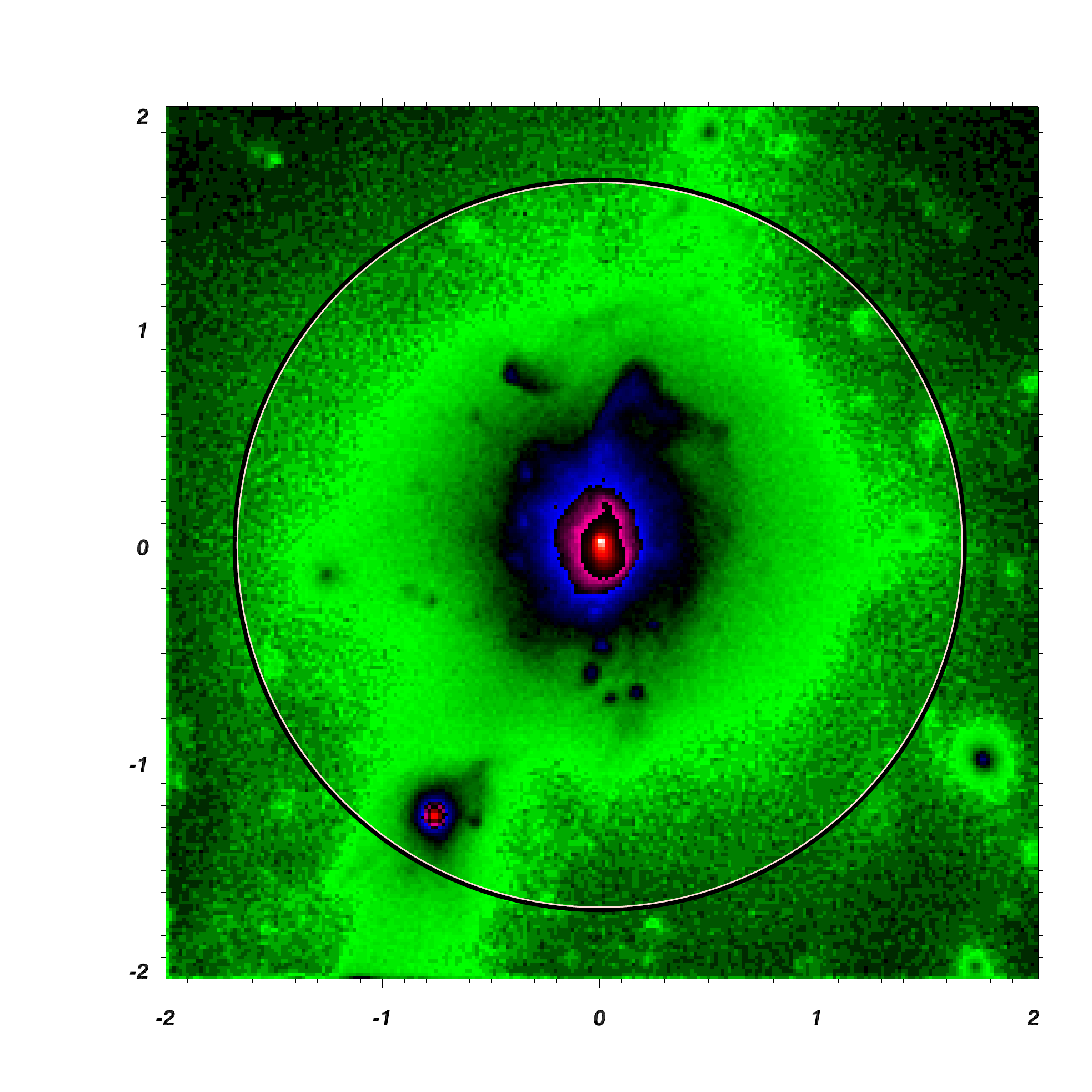}&
\includegraphics[width=50mm]{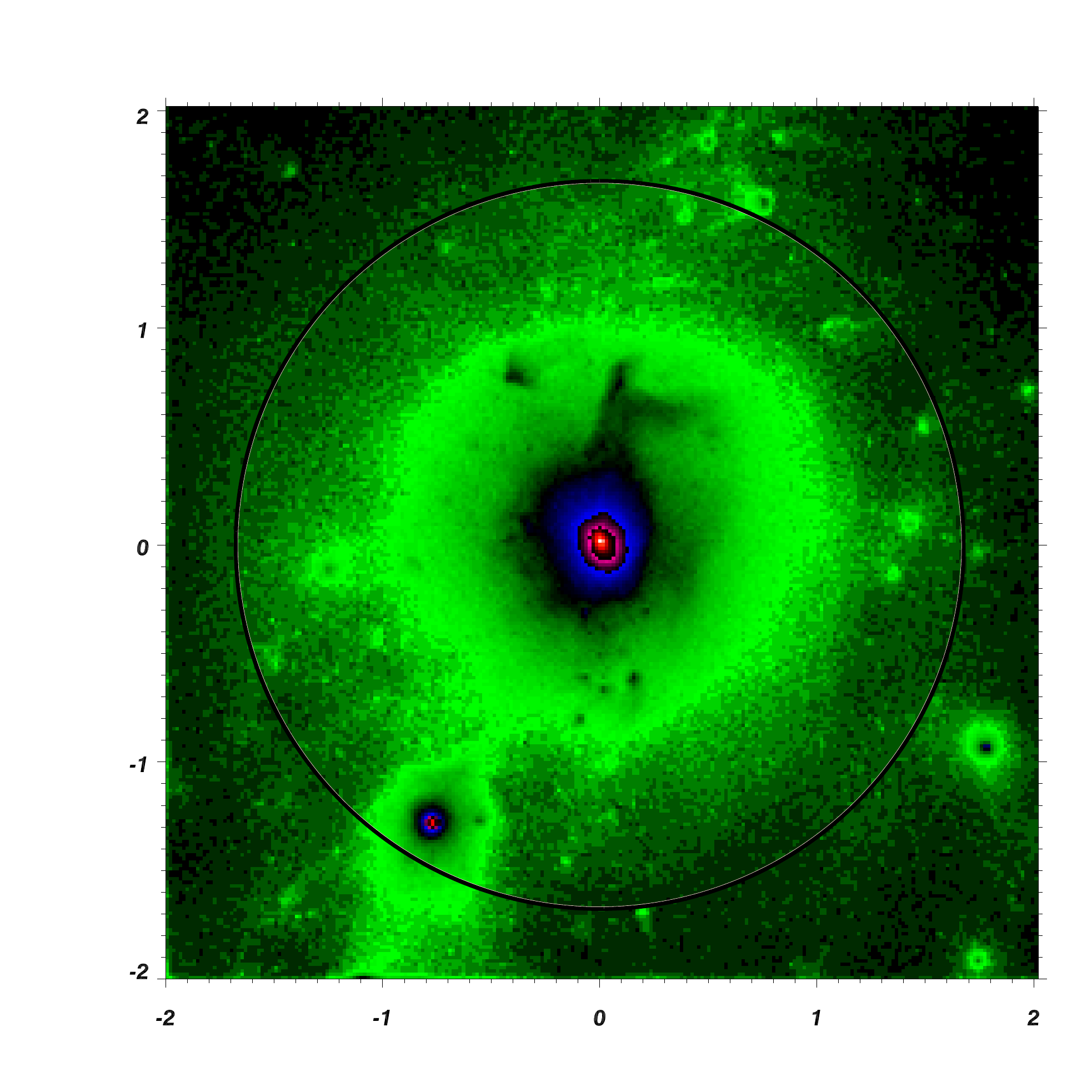} \\
G3-OWLS & G2-X & Hydra \\
\end{tabular}
\includegraphics[width=100mm]{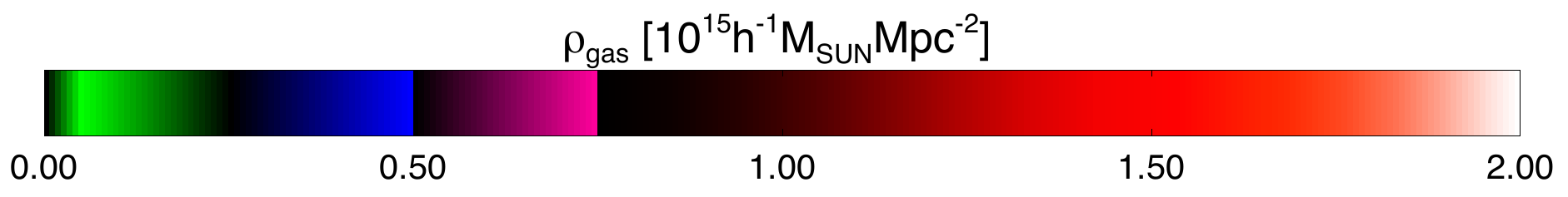} 
\caption{Projected density of the gas halo at $z=0$ for each simulation as indicated. The box is 2$h^{-1}$Mpc on a side. The white circle indicates $M_{200}^{crit}$ for the halo, the black circle shows the same but for the G3-MUSIC simulation. }
\label{fig:Non-cooling_visual_z0}
\end{figure*}

Thumbnail images of the gas
density distribution for each of the methods at $z = 0$ are given in
Figure~\ref{fig:Non-cooling_visual_z0}. We see a dramatic variation in
the central concentration of the gas, with some methods having
significantly larger extended nuclear gas regions.

\begin{figure*}
\includegraphics[width=0.7\textwidth]{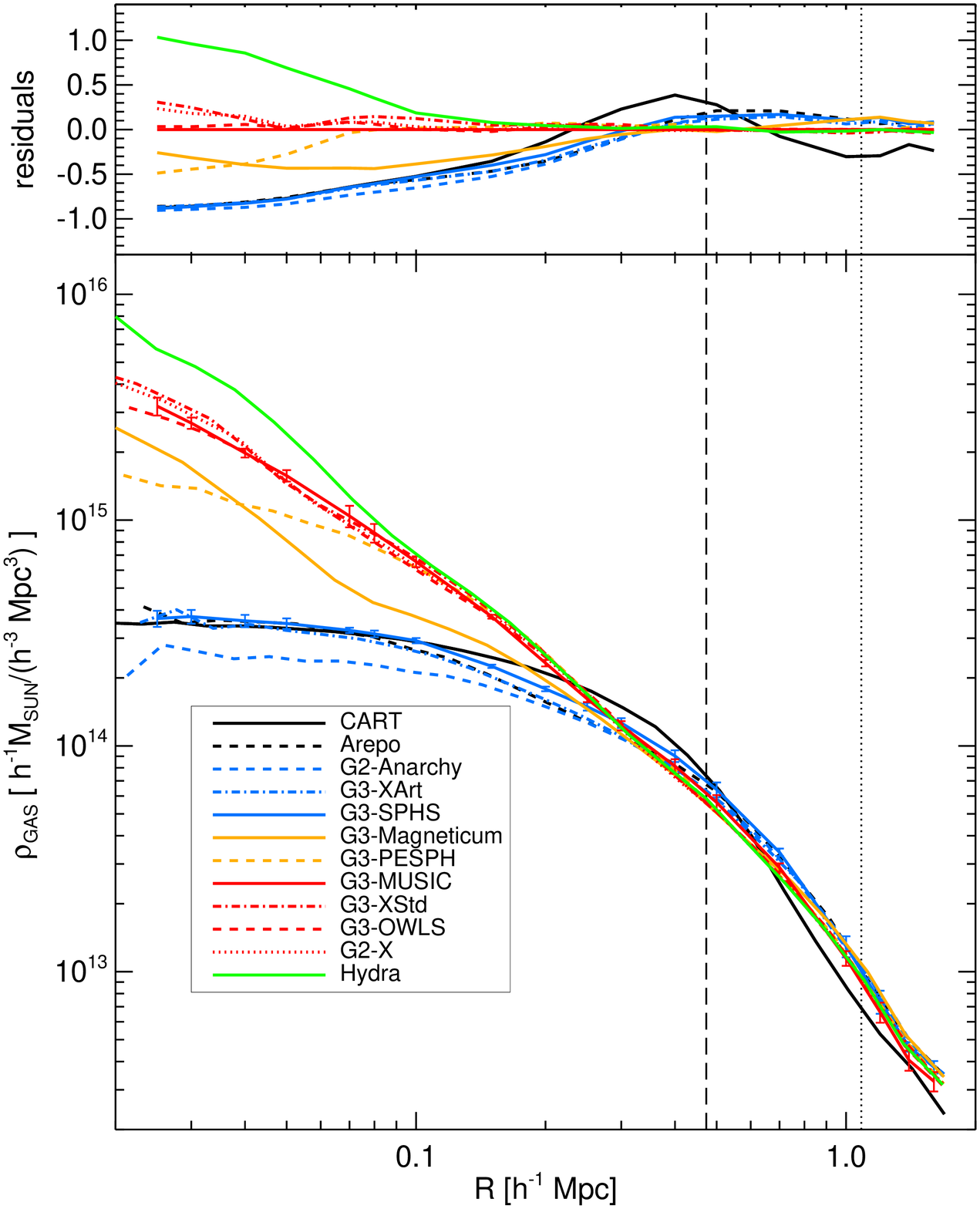}
\caption{Radial gas density profiles at $z=0$ for each simulation as indicated (bottom panel) and difference in radial gas density profiles at $z=0$ between each simulation and the reference G3-MUSIC simulation . The vertical dashed line corresponds to $R_{2500}$ and the dotted line to $R_{500}$ of the reference G3-MUSIC values. The error bars on G3-SPHS (black) and G3-MUSIC (red) are calculated from the scatter between snapshots averaged over 0.27 Gyrs. The data are cut off when the radius goes below the softening scale of the code. }
\label{fig:Non-cooling_gas_mass_profile_z0}
\end{figure*}

This trend is born out by the radial gas density profiles given in
Figure~\ref{fig:Non-cooling_gas_mass_profile_z0}. We see that the
radial gas density is significantly more extended for CART and
\arepo\ when compared to the traditional SPH schemes employed by some
SPH codes such as \hydra\ and G2-X.
SPH codes that implement artificial diffusion are quite close to CART and AREPO. Lagrangian methods with entropy mixing (including AREPO which is Lagrangian in spirit) are always very close to each other, while CART produces a gas density profile that is shallower in the central regions and steeper at larger radii.  

The difference in the radial gas density compared to the fiducial
G3-MUSIC run are shown in the top panel of 
Figure~\ref{fig:Non-cooling_gas_mass_profile_z0}. All the residuals are calculated to
the density profile of the reference G3-MUSIC simulation (and we adopt this definition for all the radiative profiles shown from now on). At $z=0$ the
lowest central densities are an order of magnitude lower than in the
G3-MUSIC simulation while the highest central densities are
around 5 times larger. i.e. the variation in the central gas density
across our runs is nearly two orders of magnitude. The scatter becomes more moderate when considering the outer region of the cluster, 
not exceeding 20 per cent at radii larger than $R_{2500}^{crit}$.

\begin{figure*}
\includegraphics[width=0.7\textwidth]{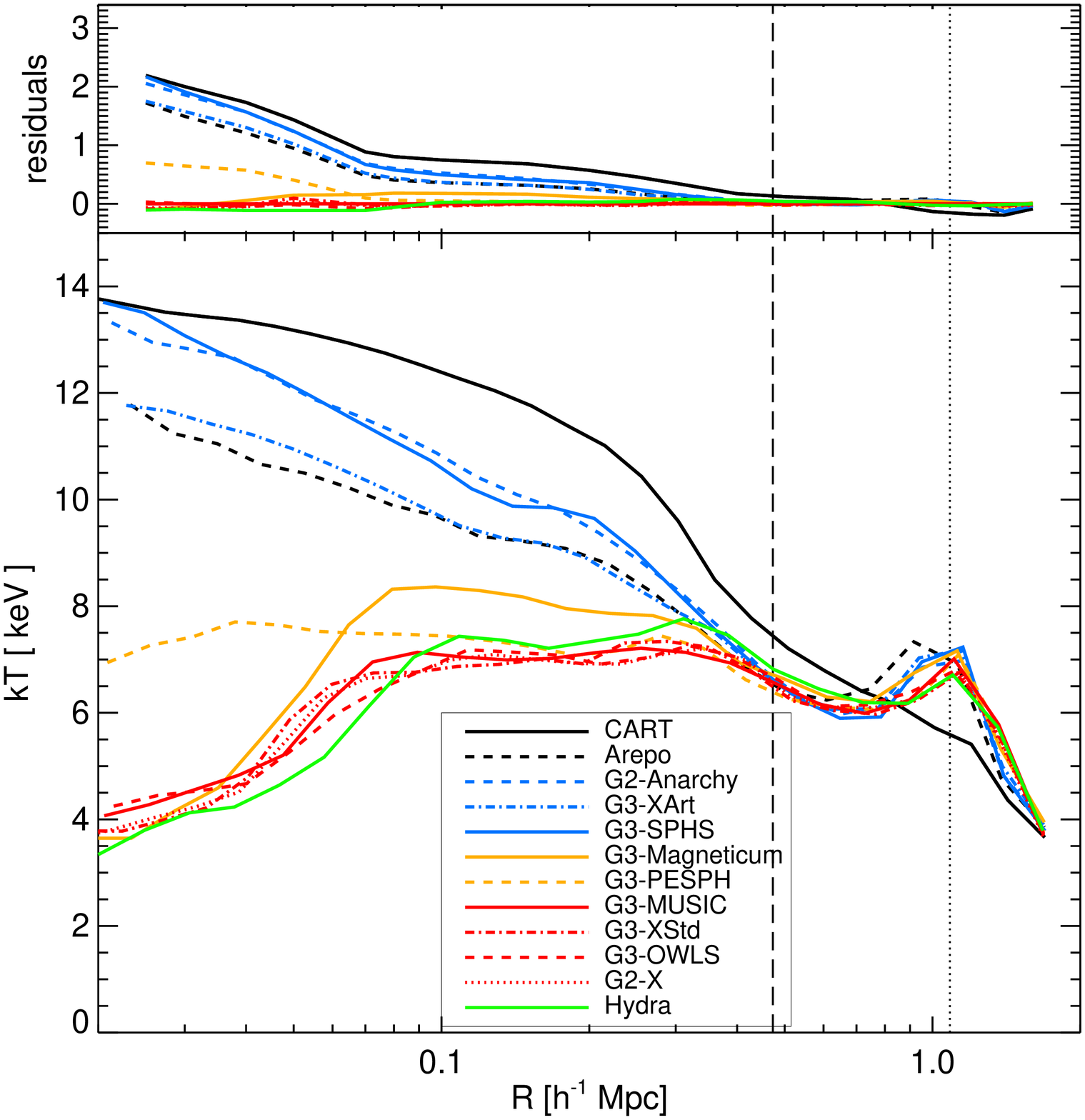}
\caption{Radial temperature profile at $z=0$ for each simulation as indicated and difference between each simulation and the reference G3-MUSIC simulation. The vertical dashed line corresponds to $R_{2500}$ and the vertical dotted line to $R_{500}$ of the reference G3-MUSIC values.}
\label{fig:temperature}
\end{figure*}

We next show the radial temperature profiles for all the simulations
in Figure~\ref{fig:temperature}. We use the mass-weighted temperature, defined as:
\begin{equation}
T_{mw} = \frac{\sum_i T_im_i}{\sum_im_i}
\end{equation}\label{eq:Tmw}
where $m_i$ and $T_i$ are the mass and the electronic temperature of the gas particles .
The central temperatures vary by more
than a factor of 3, with a group of methods displaying a central
temperature inversion with inner temperatures around half the peak
value of 7-8 keV. In contrast, some codes display a monotonically rising
temperature profile as the radius falls with a peak temperature up to 14 keV at the
very centre. This group of codes consists of those with the most extended
radial gas density profiles. Also in this case CART results are different from AREPO and SPH with thermal diffusion.
It is interesting that the jump in temperature seen at $\sim$1 $h^{-1}$Mpc for all Lagrangian methods is not seen by CART, probably because in this case the substructure that causes this bump is located in a slightly different halo position due to the differences in the merger phase.
At radii larger than $R_{2500}^{crit}$ the scatter is significantly more moderate, and the residuals appear to be a factor of 2 smaller than in Figure 17 of SB99 in the same cluster regions.

\begin{figure*}
\includegraphics[width=\textwidth]{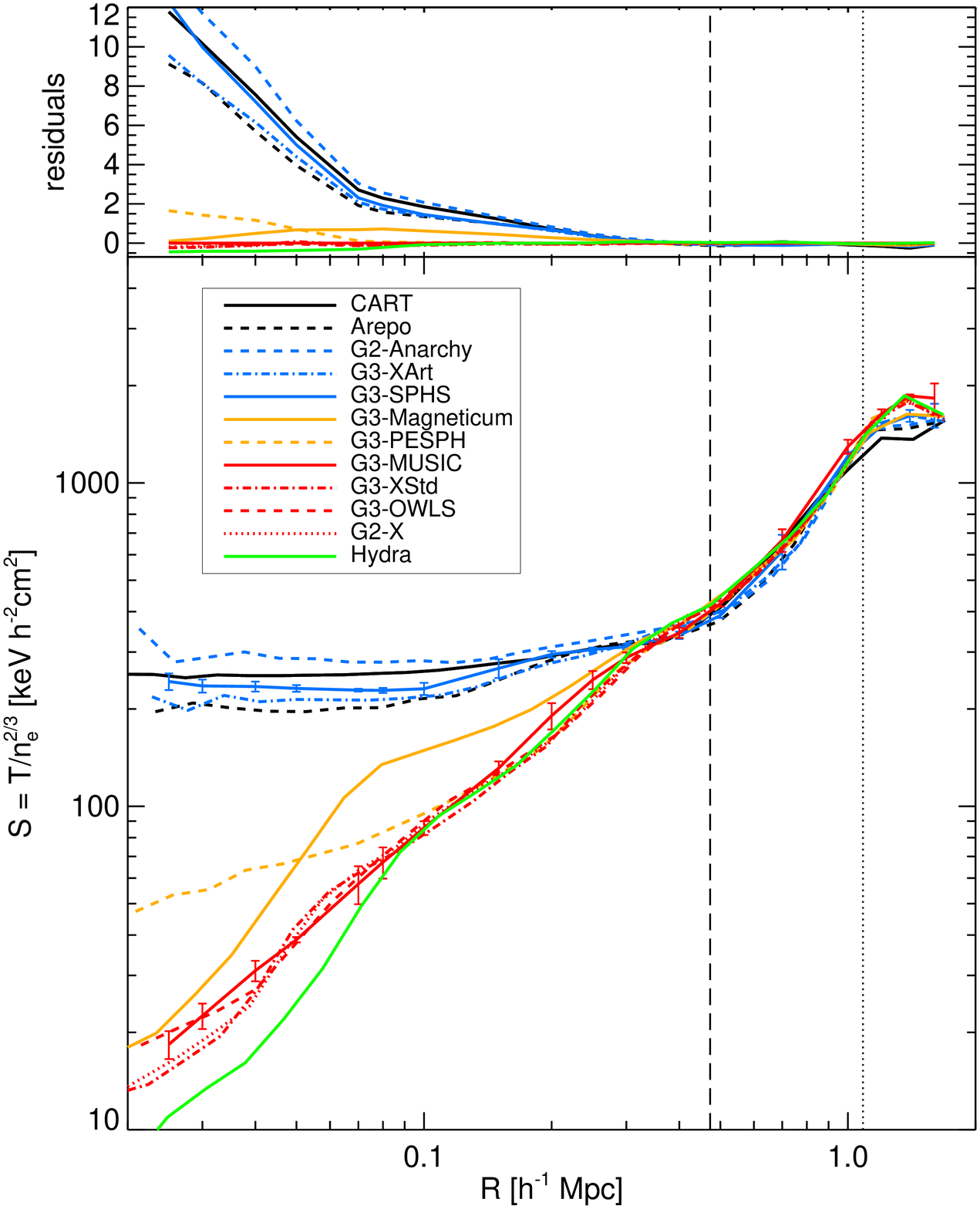}
\caption{Radial entropy at $z=0$ (bottom panel) for each simulation as indicated and difference between each simulation and the reference G3-MUSIC simulation (top panel). The dashed line corresponds to $R_{2500}$ and the dotted line to $R_{500}$ of the reference G3-MUSIC values. The error bars on G3-SPHS (black) and G3-MUSIC (red) are calculated from the scatter between snapshots averaged over 0.27 Gyrs.}
\label{fig:entropy}
\end{figure*}

Finally we show the radial gas entropy profiles for all the codes in
Figure~\ref{fig:entropy}. We adopt the definition of entropy commonly
used in the literature by works on X-ray observations:
\begin{equation}
S(R) = \frac{T_{gas}(R)}{n_e^{2/3}(R)}
\end{equation}
where $n_e$ is the number density of free electrons of the gas.
This displays the now classic split between grid-based codes and
traditional SPH methods, such as \hydra\ and G2-X, which show a
falling inner entropy as the radius is decreased all the way into the
very centre. This is completely consistent with the inner temperature
inversion and high central density. Conversely, the grid-based codes
such as CART and \arepo\ display the well known flat inner entropy
cores that result from rising inner temperature profiles and extended
gas densities. However, we see that in between these extremes we have
a full range of entropy profiles that depend on the specific SPH
implementation employed. Differently from what was shown in SB99 (see their Figure 18) 
modern, sophisticated SPH codes which employ
mixing are now capable of recovering entropy profiles that lie anywhere
between the core-less traditional SPH schemes and the cored profiles of
the grid-based approaches depending upon the precise nature of the
scheme and the amount of mixing employed. We highlight that modern SPH codes such as
G3-SPHS, G2-Anarchy and G3-XArt are able to recover the same flat entropy core observed
for CART and \arepo\, with a scatter smaller than 20 per cent, even in the inner cluster regions.
G3-PESPH and G3-Magneticum, which have artificial viscosity switch but different artificial conductivity
with respect to the other modern SPH codes,
show and intermediate behavior between standard and modern SPH codes.

\subsection{Other quantities in the non-radiative simulations} 
\begin{figure*}
\includegraphics[width=0.6\textwidth]{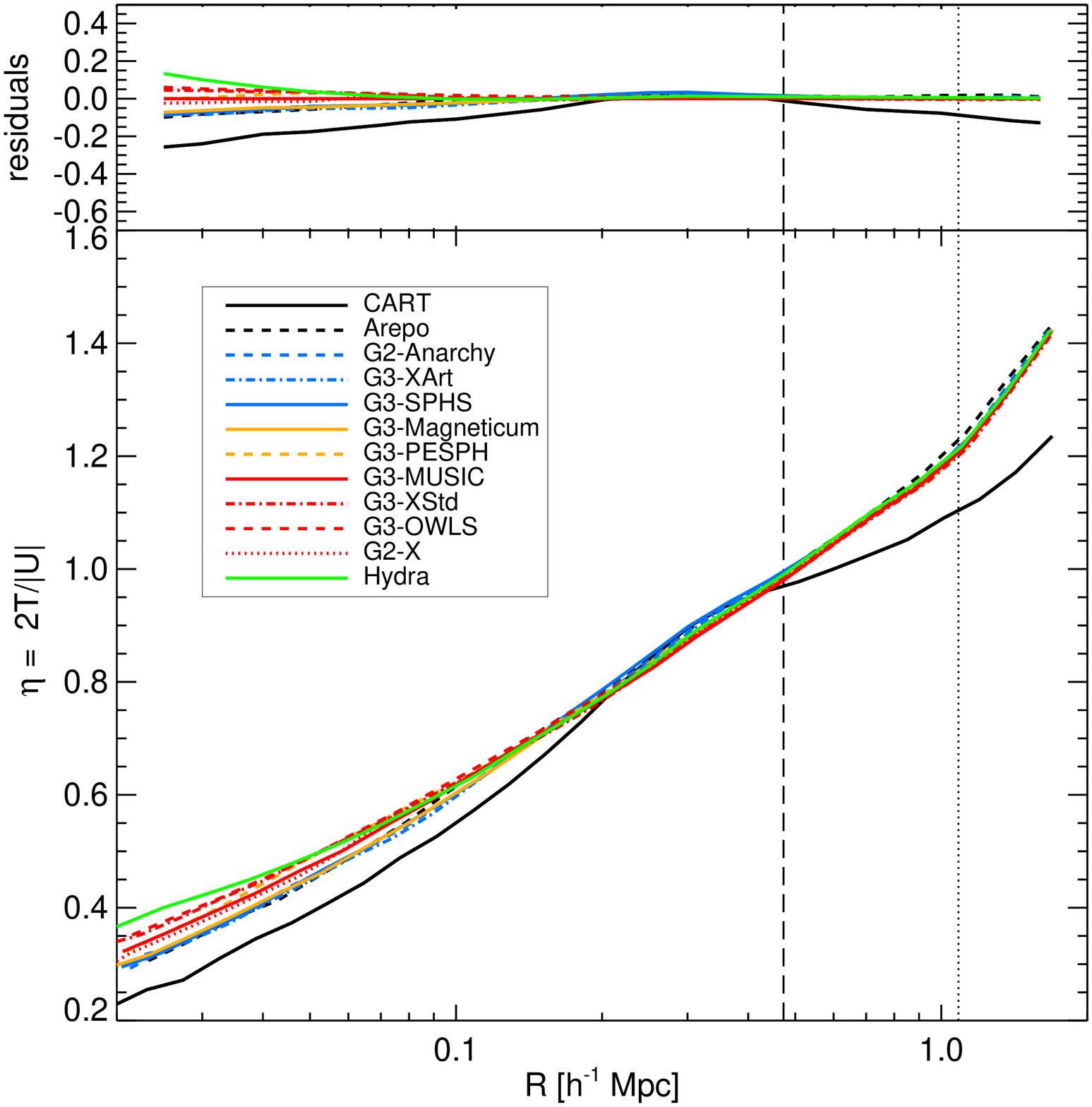}
\caption{The ratio of kinetic to thermal energy in the gas, $\eta$, measured radially at $z=0$ for each simulation as indicated (top) and difference between each simulation and the reference G3-MUSIC simulation (bottom). Dashed line corresponds to $R_{2500}$ and dotted line to $R_{500}$ of the reference G3-MUSIC values.}
\label{fig:eta}
\end{figure*}

It is important to note that the differences in radial gas density,
temperature and entropy evidenced above are not driven by code issues
such as poor thermalization or large scale flows. In
Figure~\ref{fig:eta} we show the ratio of gas thermal $U$ to kinetic
energy $K$ at $z=0$:
\begin{equation}
\eta = \frac{2K}{\mid U \mid}
\end{equation}
radial profile of all the
simulations. All the methods agree closely on the value of $\eta$ as a
function of halo radius and none display any evidence of poor
thermalization. Interestingly, we note that CART is the most efficient in converting kinetic energy into thermal energy (though also the different merger
phase may contribute to this offset).
Even considering the difference between CART and the other codes, the scatter is always below 20 per cent.

\begin{figure*}
\includegraphics[width=0.6\textwidth]{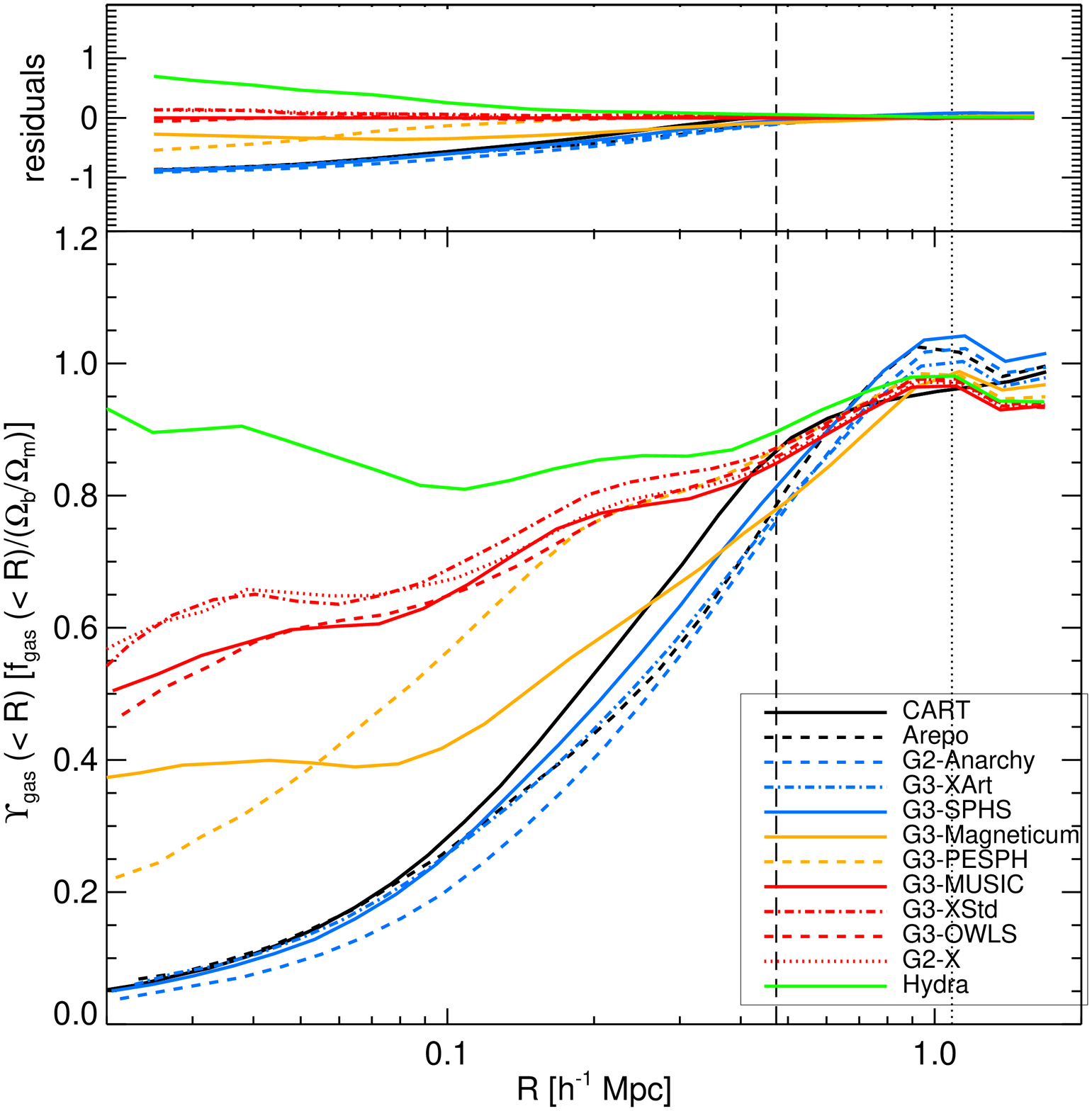}
\caption{Radial gas fraction at $z=0$ relative to the cosmic value for each simulation as indicated (top) and difference between each simulation and the reference G3-MUSIC simulation (bottom). The dashed vertical line corresponds to $R_{2500}$ and dotted vertical line to $R_{500}$ of the reference G3-MUSIC values.}
\label{fig:gasfraction}
\end{figure*}

Given our radial dark matter density and gas density profiles we can
also calculate the radial gas fraction for all the methods.
In Figure~\ref{fig:gasfraction} we show the radial profiles of the depletion factor $\Upsilon$, defined as:
\begin{equation}
\Upsilon = \frac{M_{gas}(<R)}{M(<R)}\cdot\left(\frac{\Omega_b}{\Omega_m}\right)^{-1}
\end{equation}
The results reflects the existence
of methods that produce very centrally concentrated gas and
methods yielding an extended core with much higher average entropy.  For
those codes with an extended entropy core the gas fraction falls
significantly with decreasing radius and can reach values as low as
5 per cent of the Universal baryon fraction for these non-radiative simulations. Within 100$h^{-1}$kpc it can
fall below 25 per cent of the cosmic gas fraction in both grid-based and
modern SPH codes. The differences in the radial gas fraction reflect those detected in the gas density profiles and warn about 
using a universal calibration of the baryon depletion at $R_{2500}^{crit}$ based on simulations 
in cosmological applications of the cluster baryon fraction, especially when using only non-radiative gas. We expect these results
to be very different (and closer to observational results) when radiative physics is included.
The scatter in the outer regions of the cluster
appear by the way to be much more limited (less than 20 per cent) with respect to the results shown in Figure 13 of SB99, where
differences of up to 50 per cent between the different codes where registered even close to the virial radius.

\begin{figure*}
\includegraphics[width=0.6\textwidth]{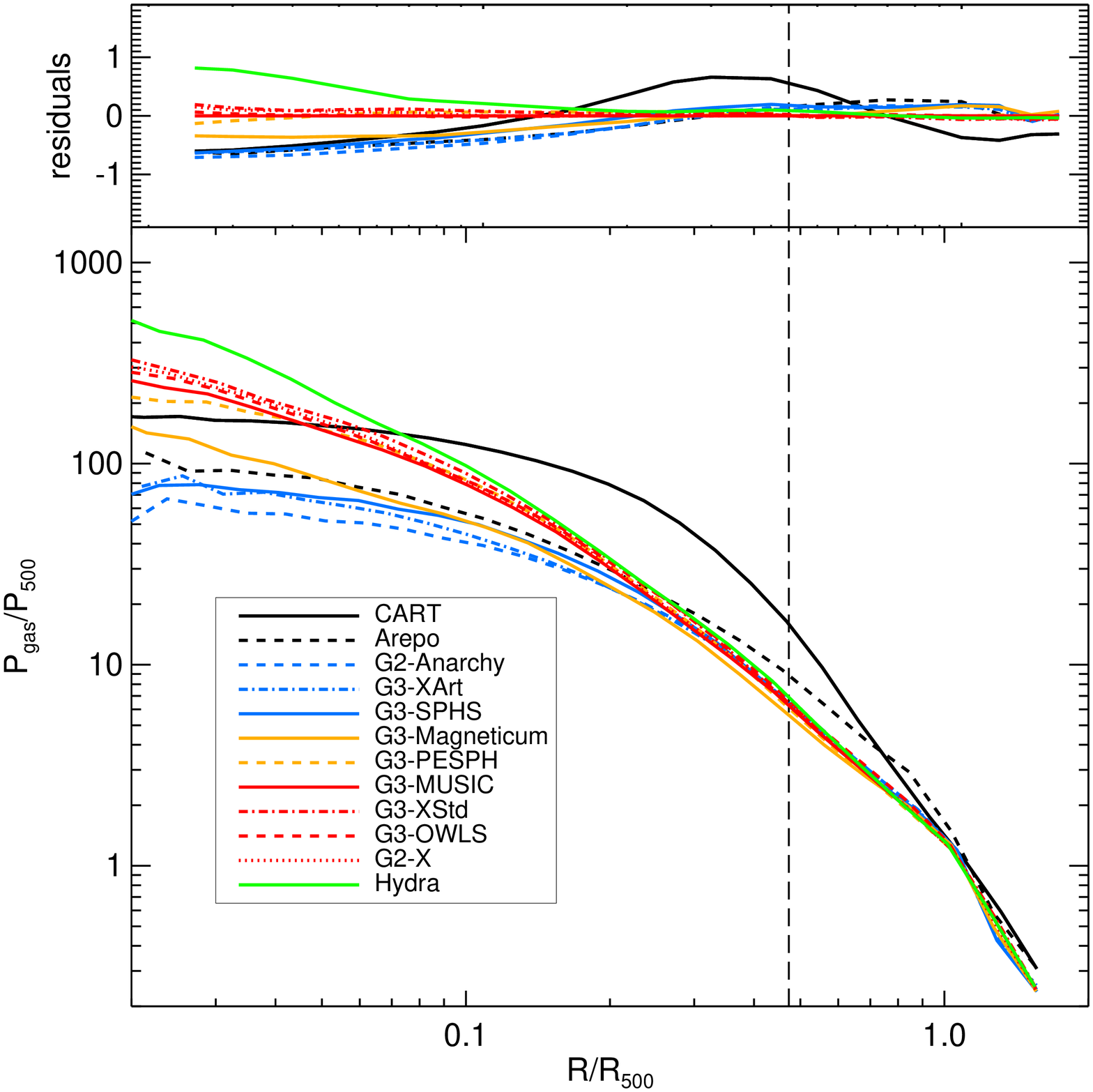}
\caption{Radial gas pressure at $z=0$ measured in each simulation as indicated (bottom panel) and difference between each simulation and the reference G3-MUSIC simulation (top panel). The pressure, as well as the radius, has been normalized to the corresponding value at $R_{500}$ for each code. The dashed vertical line corresponds to $R_{2500}$  of the reference G3-MUSIC value.}
\label{fig:pressure}
\end{figure*}

\begin{figure*}
\includegraphics[width=0.6\textwidth]{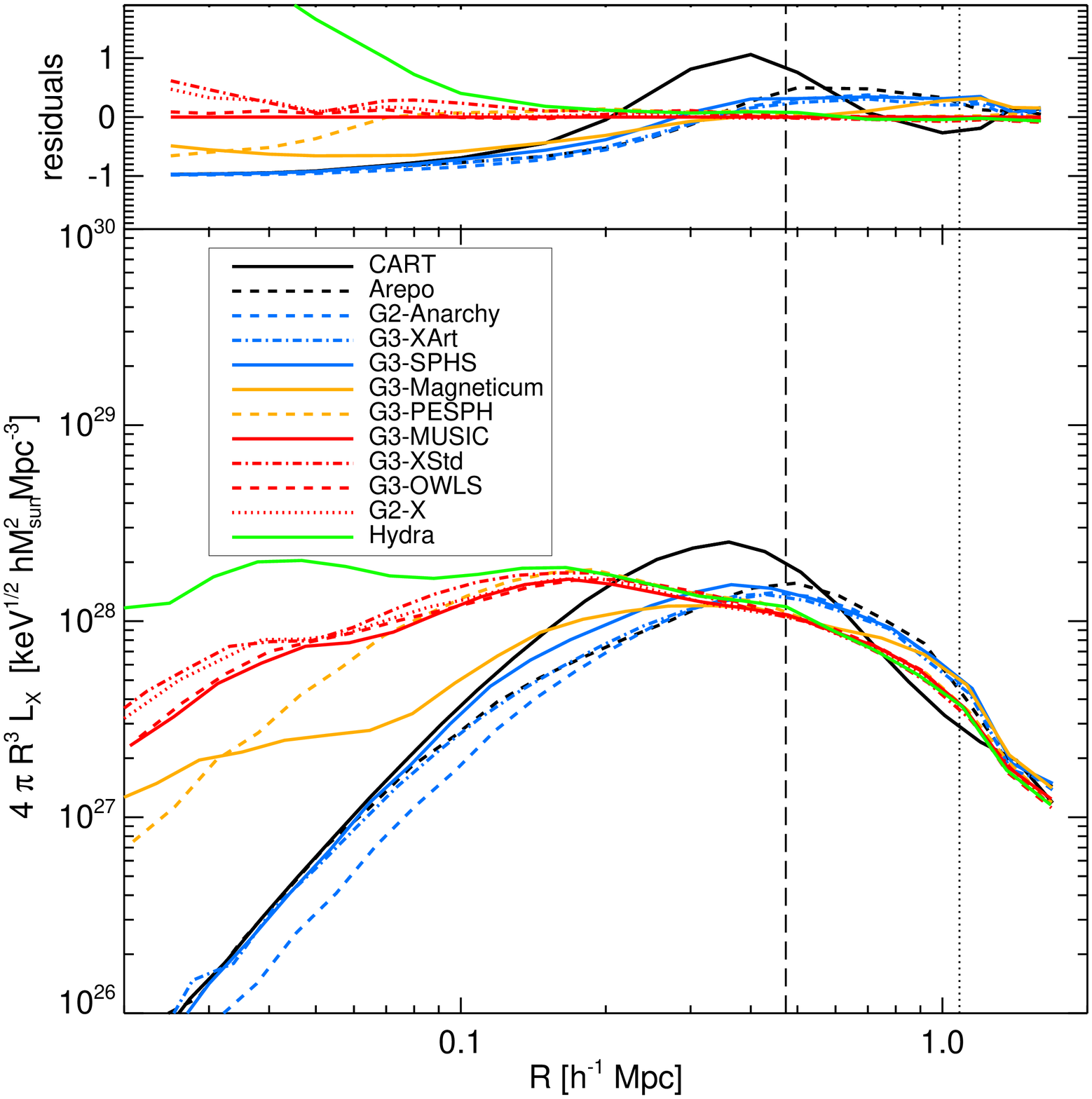}
\caption{Radial X-ray luminosity profiles at $z=0$ for each simulation as indicated (bottom panel) and difference between each simulation and the reference G3-MUSIC simulation (top panel)..The dashed vertical line corresponds to $R_{2500}$  of the reference G3-MUSIC value.}
\label{fig:xray}
\end{figure*}

We can combine our measurements of the gas density and temperature to
produce Figure~\ref{fig:pressure}, the gas pressure profiles
and Figure~\ref{fig:xray}, the X-ray emission profiles. 
We define the pressure as $P = \rho_{gas}T$ and we normalize the profiles
to the value of $P_{500}$ (the value of the pressure as calculated at $R_{500}^{crit}$) in order
to be consistent with the definition of universal pressure profile introduced by \cite{Arnaud10}.
The X-ray luminosity profile is defined as $4\pi R^3L_X$ and we approximate the X-ray luminosity density
as $L_X$ = $\rho_{gas}^2T^{1/2}$.
The variation in the gas density and temperature produce very different pressure and
X-ray emission profiles. As expected, the
pressure profiles continue to rise all the way into the centre for all
codes (as the central gas is close to hydrostatic equilibrium in all
cases). Due to the very high central density, the central X-ray
emission for \hydra\ is over two orders of magnitude higher than that
found by the grid-based and some modern SPH methods which form extended cores.
The differences observed in CART profiles, especially in Figure~\ref{fig:pressure}, can be attributed to the different merger phase.

\section{Convergence and scatter between simulations}\label{sec:scatter}
 
\subsection{Dark matter only runs}  The major cause of the discrepancy (for the \Gadget\ 
based codes at least) is the size of the base level particle mesh.
A base mesh of $256^3$ is not sufficient to resolve the interface region between the low resolution region (i.e. the base level) and the
high resolution region (i.e. the refined level) placed over the cluster; we found that a base level of at least $512^3$ is required to ensure that
the dark matter component is well aligned across the different codes. 
For the $N$-body only simulations we find excellent agreement between the density profiles of the main halo and the statistics of their subhalos (Figures \ref{fig:DM_density_profile} and \ref{fig:DM_subhalo_massfunc_z0}), once the input parameters of the various runs are matched (see Appendix). However, even the 'matched' simulations show important phase differences between simulations (Figure A1). The numerous implementations of \Gadget\ agree remarkably well, but this is perhaps not surprising given that the gravity solver is in each case built on the same foundation. CART, however, shows a marked disagreement, as a large sub halo appears to be absent. Similar results were reported recently in \citet{Power14} when comparing G3-SPHS with the \Ramses\ AMR code (see their Figure 7). It is not clear that this is either a problem or unexpected. We are, after all, modeling a chaotic system and while we can expect the {\it statistics} of the density field to agree between codes, it is 
likely unreasonable to expect the precise merger history and merger phase to agree between codes for a single object. Ideally, we would compare the halo statistics between codes for a large ensemble of cluster zoom simulations. Such a study is beyond the scope of this present work and will treated in an future paper dedicated to the study of the subhalos.

\subsection{Adding gas: AMR versus SPH} Issues such as the merger phase appear to affect the resulting dark matter properties at the $\sim$ 10 per cent level (Figures \ref{fig:DM_density_profile}-\ref{fig:DM_subhalo_massfunc_z0}). However, once adding gas, the merger phase produces a larger effect than this on the resulting gas distribution. This is because entropy in clusters is generated in shocks during collapse and mergers \citep{2009MNRAS.395..180M, Power14}. The different codes explored here broadly fall into four categories. There are the `classic' SPH methods (\Hydra; G3-MUSIC; G3-OWLS; G3-XStd; G2-X); the 'modern' SPH methods that attempt to correct for problems with mixing in classic SPH (G3-XArt; G3-SPHS; G2-Anarchy; G3-PESPH; G3-Magneticum); an Adaptive Mesh refinement method (CART) and a hybrid `moving mesh' method (\Arepo; see section \ref{sec:codes} for further details of each method and the differences between them). The AMR code CART that is the most discrepant in its dark matter distribution is also the most discrepant in in its gas distribution, particularly at large radii. This is almost certainly attributable to the different merger phase, as can be seen from Figure A1.

\subsection{The trouble with `classic' SPH} Despite the excellent agreement between the `dark matter only' runs, the {\rm classic} SPH simulations also show scatter in their central gas density, temperature and entropy that is significantly larger than the theoretical error on each (as calculated from the scatter between snapshots averaged over 0.27 Gyrs;  see the error bars marked on the plots shown in Figures \ref{fig:Non-cooling_gas_mass_profile_z0} and \ref{fig:entropy}). This is likely because in {\rm classic} SPH, low entropy particles artificially sink to the cluster centre due to the lack of mixing \citep{Power14}, amplifying the effect of small changes in merger phase. We see this in Figure \ref{fig:Non-cooling_visual_z0}.

\subsection{Modern SPH and \Arepo} Of considerable interest are the differences between the {\rm modern} SPH flavors (G3-XArt; G3-SPHS; G2-Anarchy; G3-PESPH; G3-Magneticum) and the moving mesh code \Arepo.
There is excellent agreement between the dark matter distributions across the runs, which allows us to isolate the effect of the hydrodynamics solver (Figure A1). For the gas distributions shown in Figure \ref{fig:Non-cooling_gas_mass_profile_z0}, \Arepo\, G3-XArt and G3-SPHS are in excellent agreement with one another, agreeing almost perfectly within our estimated theoretical error, while G2-Anarchy seems to be the outlier. In Figure \ref{fig:temperature} the temperature profiles for \Arepo\; and G3-XArt are very closely matched but look to have an offset from G3-SPHS and G2-Anarchy, which are very close to one other. In Figure \ref{fig:entropy} the radial entropy profiles of G3-SPHS, G3-XArt and \Arepo\; are again very close to one another while G2-Anarchy is more discrepant, though closer to CART. The origin of these differences is yet to be explained, although we note that it cannot be attributed to different merger phases and must result from the hydrodynamics solver. In the case of G2-Anarchy a possible cause of discrepancy may be the choice of the kernel (using a C4 kernel with 200 neighbors gives slightly different values for the central entropy). \citet{Power14} showed that at the resolution of the simulations in this paper, G3-SPHS is numerically converged. It would be interesting to see if the differences between G3-SPHS/G3-XArt/Arepo and G2-Anarchy remain with increasing resolution. We defer such a resolution study to future work, whose intent will be to narrow down these numerical uncertainties.

In the Arepo simulations we have used the total energy as a conserved quantity in the Godunov scheme, which is the usual choice in finite volume codes and has been used in most recent Arepo studies (as, e.g., in \citealt{Vogelsberger14}). As discussed in \cite{Springel10}, using an entropy-energy formalism results in smaller entropy cores and higher central gas densities somewhat closer to classic SPH results (similar results are also shown by GIZMO, \citealt{Hopkins14}). Following most recent Arepo studies, we have not employed the later method here due to concerns regarding the artificial suppression of weak shocks and the potentially less accurate energy conservation.

Finally, we note that the results of G3-PESPH are very different
from those of the other ÕmodernÕ SPH flavours (with the
exception of G3-Magneticum), and are more similar to those of
classic SPH simulations. A key difference is that this version
of G3-PESPH does not include any explicit conductivity or mixing,
while all the other modern variants do.  \cite{hu14} showed
that PESPH performs much better than previous versions of SPH for
surface instabilities by greatly mitigating surface tension problems,
but in areas of very strong shocks (M$\sim$1000) adding
artificial conduction provides a better match to analytic solutions. 
Insights into the behavior of G3-PESPH may be gained by considering the ÔOSPHÕ method
presented in \citet{2009arXiv0906.0774R}, and the earlier multiphase
ÕRTÕ method by \citet{2001MNRAS.323..743R}. As pointed out
by \citet{2012MNRAS.422.3037R}, the ÔRTÕ method only works well for
relatively small entropy contrasts between different fluid phases. As
the entropy contrast becomes very large, the admixture of low and
high entropy particles within the smoothing kernel creates large
pressure fluctuations that prevent mixing, as in classic SPH. This
was recognized also by  \citet{2001MNRAS.323..743R} who proposed
scaling the neighbor number with the entropy contrast to combat
this. However, this rapidly becomes prohibitively expensive in realworld
applications, which led \citet{2012MNRAS.422.3037R} to abandon
such ÕRTÕ methods in favor of dissipation switching, as proposed
by \citet{Price2008}; such switching is common to G3-XArt, G3-SPHS
and G2-Anarchy and helps to explain their similarity. The discrepancy
between G3-PESPH and the other modern SPH flavours may
also reflect the treatment of artificial viscosity Ð adopting the artificial
viscosity model suggested by \cite{Cullen10}  can produce
larger entropy gains in shocks, but the authors of G3-PESPH do not use
this because it also seems to add substantial entropy into very diffuse intergalactic
gas that may be spurious (N. Katz, priv. comm.).  In short, it is unclear how much artificial
conductivity and/or mixing is appropriate in SPH, or even whether the mesh
codes are providing the correct solution that all SPH codes should be trying to emulate.
Nonetheless, consistency with mesh codes appears to require modern SPH using
conduction, mixing, and/or a higher-order dissipation switch.

The discrepancy between G3-PESPH and the other modern SPH flavours may also reflect the treatment of artificial viscosity -- adopting the artificial viscosity model suggested by \cite{Cullen10} should lead to a better agreement with mesh-based codes. 

\section{Summary \& Conclusions}\label{sec:conclusion}

We have investigated the performance of 12 modern astrophysical simulation
codes -- CART, \hydra, \arepo\ and 9 versions of \gadget\ with different SPH
implementations -- by carrying out cosmological zoom simulations of a single
massive galaxy cluster. Our goal was to assess the consistency of the different
codes in reproducing the spatial and thermodynamical structure of dark matter
and non-radiative gas in the cluster.

As our initial step, we ran dark matter only versions of the simulations with each
code using its preferred set of numerical parameters (e.g. time step accuracy,
gravitational softening, dimension of the particle mesh), and examined the spherically
averaged mass density profile and the spatial distribution of substructures. We found
good consistency between the density profiles recovered by the codes at approximately
the 10 per cent level, while there were small variations in the positions of substructures.
When these simulations were re-run with a common set of numerical parameters, we found
that these small variations could be suppressed (essentially entirely, in the case of the
\gadget\ codes).

By adopting this common parameter set, we were able to isolate those differences
between the results of the hydrodynamical simulations that arise only from the
choice of hydrodynamical solver, rather than from the complex interplay of the hydrodynamical
and gravity solvers. Interestingly, we found that the resulting gas density profiles varied
substantially amongst the codes. Our key findings can be summarized as follows:

\begin{itemize}
\item Some codes, essentially the oldest, with {\rm classic} SPH implementations, exhibit
  continually falling inner entropy profiles, without any evidence of an entropy core. This
  is because these codes, particularly \hydra\ ,were carefully designed to be entropy
  conserving with very low levels of mixing. This lack of mixing preserves low-entropy gas
  particles at the centers of all objects, including subhalos, which survive until late times.
  As the cluster relaxes, these particles sink to the centre of the radial density profile,
  decreasing the central entropy.
\item In contrast, the grid-based codes CART and \arepo\ produce extended cores
  with a large constant entropy core. In these mesh based codes mixing of entropy arises as a
  consequence of the numerical diffusion associated with the Riemann solver: they naturally mix
  entropy between gas elements, essentially eliminating the very low entropy material.
\item Modern SPH codes such as \textsc{G3-ANARCHY}, \textsc{G3-SPHS} and \textsc{G3-XArt},
  which have dissipative switches and new kernels, can bridge the gap between the classic SPH
  codes and grid based codes, and produce entropy cores that are indistinguishable from those of
  the grid-based codes. 
\end{itemize}

Our results confirm that the discrepancies between grid-based codes and SPH codes in describing
the radial entropy profile of simulated clusters, identified by the Santa Barbara comparison
project presented in \cite{Frenk99}, can be overcome by {\rm modern} SPH codes. Importantly, all
the codes employed in this work succeed in recovering the global properties and most of the radial
profiles of a simulated large galaxy cluster with much greater accuracy and significantly smaller
scatter than those presented in \cite{Frenk99}; this highlights the enormous strides in the
development of astrophysical hydrodynamical simulation codes over the last decade.

This work constitutes the first in a series of papers in which we examine in detail the predictions
of modern astrophysical hydrodynamical simulation codes. The next paper in this series will focus on 
simulations of the same galaxy cluster, now modeled with a variety of galaxy formation processes
including cooling, star formation, supernovae, and feedback from active galactic nuclei (AGN). This
will allow us to establish how radiative processes affect the entropy cores of simulated clusters.
Subsequent papers will look at the recovery of cluster properties such as X-ray temperature and
Sunyaev-Zel'dovich profiles; gravitational lensing; and cluster outskirts and hydrostatic mass bias;
all of which will add to our understanding of how consistently the results of different codes can
inform our understanding of galaxy cluster properties.

\section*{Acknowledgments}
The authors would like to express special thanks to the Instituto de Fisica Teorica (IFT-UAM/CSIC in Madrid) for its hospitality and support, via the Centro de Excelencia Severo Ochoa Program under Grant No. SEV-2012-0249, during the three week workshop "nIFTy Cosmology" where this work developed. We further acknowledge the financial support of the University of Western 2014 Australia Research Collaboration Award for "Fast Approximate Synthetic Universes for the SKA", the ARC Centre of Excellence for All Sky Astrophysics (CAASTRO) grant number CE110001020, and the two ARC Discovery Projects DP130100117 and DP140100198. We also recognize support from the Universidad Autonoma de Madrid (UAM) for the workshop infrastructure.

AK is supported by the {\it Ministerio de Econom\'ia y Competitividad} (MINECO) in Spain through grant AYA2012-31101 as well as the Consolider-Ingenio 2010 Programme of the {\it Spanish Ministerio de Ciencia e Innovaci\'on} (MICINN) under grant MultiDark CSD2009-00064. He also acknowledges support from the {\it Australian Research Council} (ARC) grants DP130100117 and DP140100198. He further thanks France Gall for le coeur qui jazze. CP acknowledges support of the Australian Research Council (ARC) through Future Fellowship FT130100041 and Discovery Project DP140100198. WC and CP acknowledge support of ARC DP130100117. EP acknowledges support by the ERC grant ``The Emergence of Structure during the epoch of Reionization''. STK acknowledges support from STFC through grant ST/L000768/1. RJT acknowledges support via a Discovery Grant from NSERC and the Canada Research Chairs program. Simulations were run on the CFI-NSRIT funded Saint Mary's Computational Astrophysics Laboratory. SB \& GM acknowledge support from the PRIN-MIUR 2012 Grant "The Evolution of Cosmic Baryons" funded by the Italian Minister of University and Research, by the PRIN-INAF 2012 Grant "Multi-scale Simulations of Cosmic Structures", by the INFN INDARK Grant and by the "Consorzio per la Fisica di Trieste". IGM acknowledges support from a STFC Advanced Fellowship. DN, KN, and EL are supported in part by NSF AST-1009811, NASA ATP NNX11AE07G, NASA Chandra grants GO213004B and TM4-15007X, the Research Corporation, and by the facilities and staff of the Yale University Faculty of Arts and Sciences High Performance Computing Center. PJE is supported by the SSimPL program and the Sydney Institute for Astronomy (SIfA), DP130100117. JIR acknowledges support from SNF grant PP00P2 128540/1. CDV acknowledges financial support from the Spanish Ministry of Economy and Competitiveness (MINECO) through the 2011 Severo Ochoa Program MINECO SEV-2011-0187 and the AYA2013-46886-P grant. AMB is supported by the DFG Research Unit 1254 'Magnetisation of Interstellar and Intergalactic Media' and by the DFG Cluster of Excellence 'Origin and Structure of the Universe'. RDAN acknowledges the support received from the Jim Buckee Fellowship. The AREPO simulations were performed with resources awarded through STFCs DiRAC initiative. The authors thank Volger Springel for helpful discussions and for making AREPO and the original GADGET version available for this project.

The authors contributed to this paper in the following ways: FS, GY, FRP, AK, CP, STK \& WC formed the core team that organized and analyzed the data, made the plots and wrote the paper. AK, GY \& FRP organized the nIFTy workshop at which this program was completed. GY supplied the initial conditions. PJE assisted with the analysis. All the other authors, as listed in Section~\ref{sec:codes} performed the simulations using their codes. All authors had the opportunity to proof read and comment on the paper. 

The simulation used for this paper has been run on Marenostrum supercomputer and is publicly available at the MUSIC website\footnote{\href{http://music.ft.uam.es}{http://music.ft.uam.es}}.

This research has made use of NASA's Astrophysics Data System (ADS) and the arXiv preprint server.

\bibliographystyle{mn2e}

\section{Dark matter alignment} 

\begin{table*}
\caption{Numerical parameters used for the \arepo \ and \gadget \ runs: accuracy of the time step (ETIA), time step displacement factor (MRDF), minimum (MINT) and maximum (MAXT) time step, tolerance on the force accuracy (ETFA), accuracy of the tree algorithm (ETT), Courant factor (CFAC) and double precision (DP, DF). }
\label{tab:gadgetparams}
\begin{center}
\begin{tabular}{llllllllll}
\hline
Code name & ETIA & MRDF & MINT & MAXT & ETFA & ETT & CFAC & DP & DF \\
\hline
\arepo 					& 0.025 & 0.125 & 0 & 0.01 & 0.0025 & 0.6 &  0.3 & Y & Y\\
\gadgettwox				& 0.02 & 0.25 & $10^{-7}$ & 0.025 & 0.0025 & 0.3 & 0.15 & Y & Y\\
\gadgetmagneticum			& 0.05 & 0.25 & 0 & 0.05 & 0.005 & 0.45 & 0.15 & Y & Y\\
\gadgetmusic				& 0.01 & 0.5 & 0 & 0.01 & 0.01 & 0.4 & 0.15 & Y & Y\\
\gadgetowls				& 0.025 & 0.25 & $10^{-10}$ & 0.025 & 0.005 & 0.6 & 0.15 & Y & N\\
\gadgetsphs				& 0.03 & 0.5 & 0 & 0.02 & 0.005 & 0.5 & 0.4 & N & N\\
\gadgetx					& 0.01 & 0.5 & 0 & 0.01 & 0.01 & 0.45 & 0.15 & N & N\\
\gadgetwindy				& 0.05 & 0.25 & $10^{-7}$ & 0.05 & 0.005 & 0.4 & 0.15 & Y & Y\\
\gadgetanarchy				& 0.01 & 0.125 & 0.0 & 0.01 & 0.025 & 0.3 & 0.3 & Y & Y\\
\hline
Common parameter set			& 0.01 & 0.125 & 0.0 & 0.01 & 0.025 & 0.3 & 0.15 & Y & Y\\
\end{tabular}
\end{center}
\end{table*}

\begin{figure*}
\centering
\begin{tabular}{ccc}
\includegraphics[width=54mm]{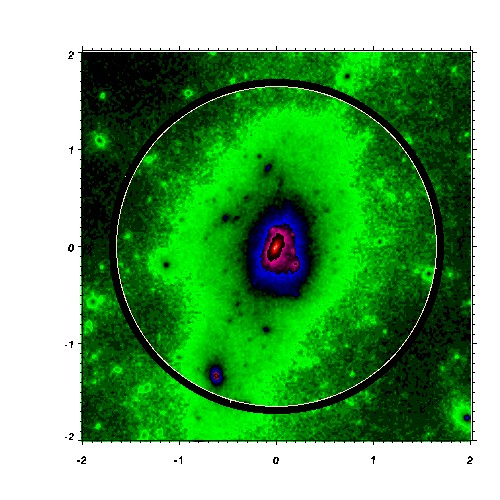}&
\includegraphics[width=54mm]{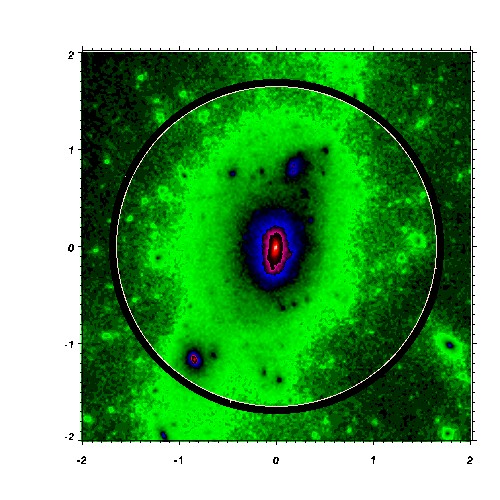}&
\includegraphics[width=54mm]{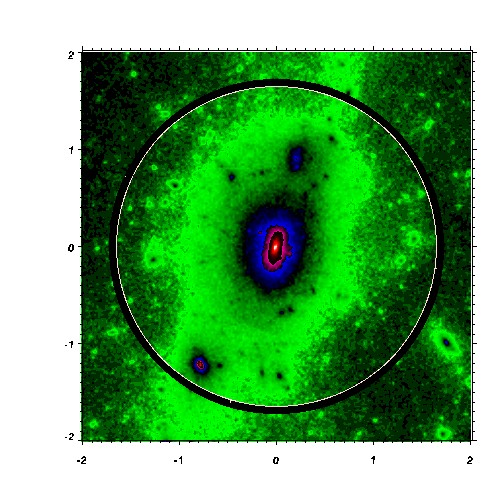}\\
CART & Arepo & G2-Anarchy\\
\includegraphics[width=54mm]{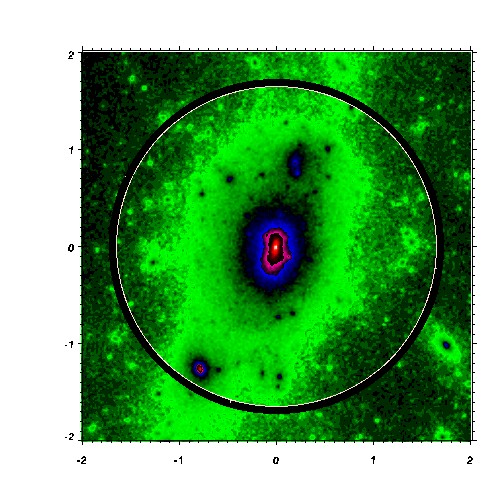}&
\includegraphics[width=54mm]{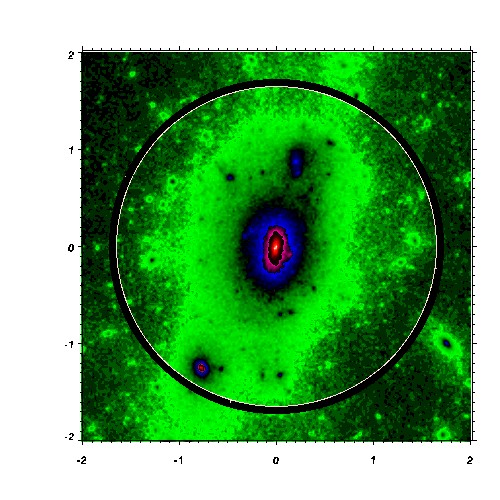}&
\includegraphics[width=54mm]{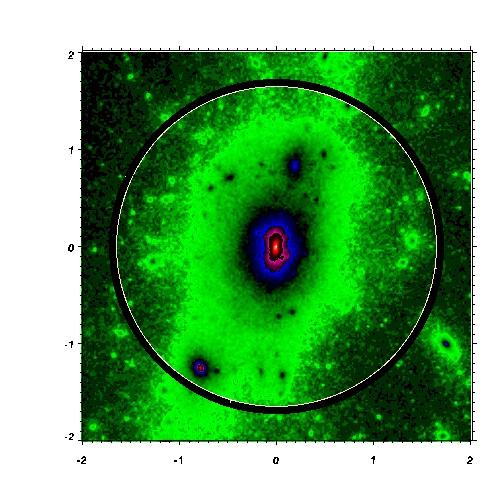}\\
G3-XArt & G3-SPHS & G3-Magneticum\\
\includegraphics[width=54mm]{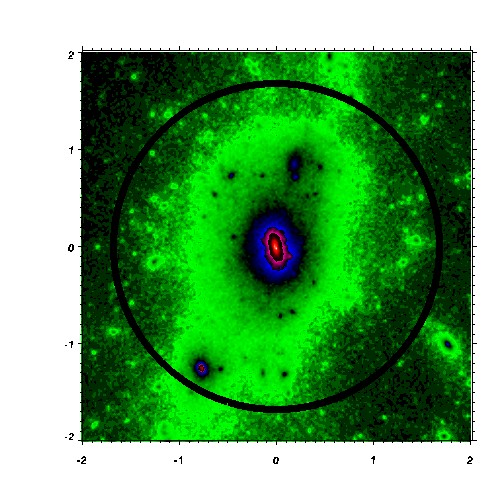}&
\includegraphics[width=54mm]{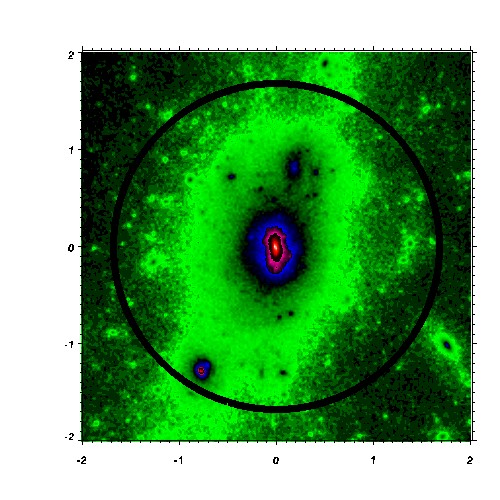}&
\includegraphics[width=54mm]{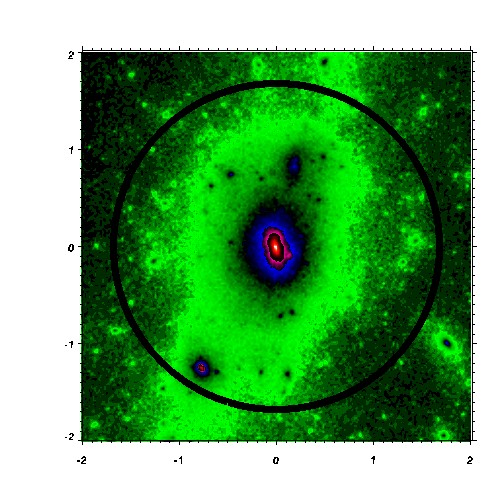}\\
G3-PESPH & G3-MUSIC & G3-XStd\\
\includegraphics[width=54mm]{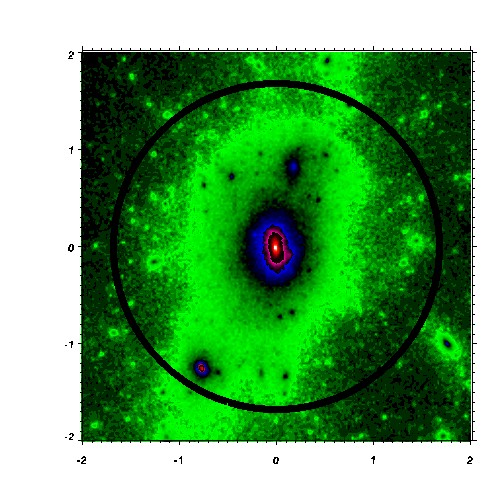}&
\includegraphics[width=54mm]{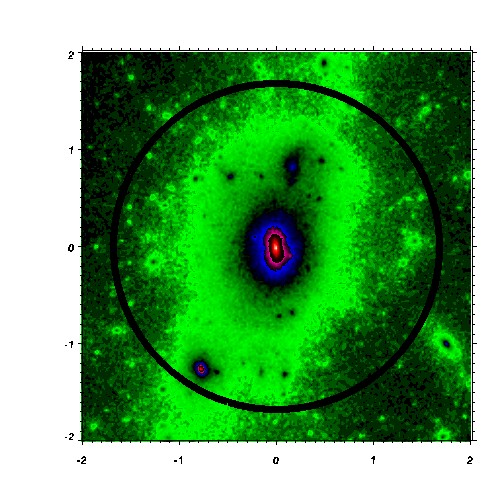}&
\includegraphics[width=54mm]{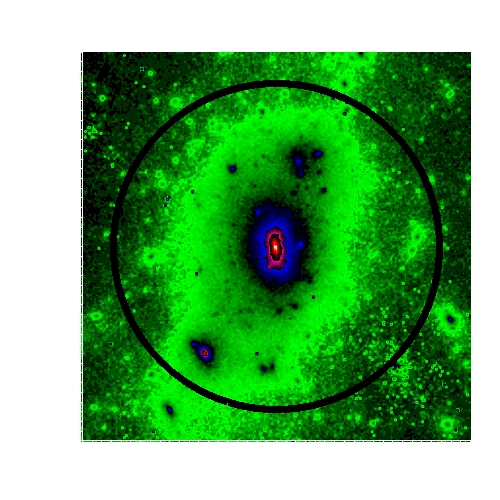}\\
G3-OWLS & G2-X & Hydra\\
\end{tabular}
\includegraphics[width=100mm]{plots/new/dm/maps/z0/density_bar} 
\caption{Projected density of the dark matter halo in the non-radiative simulations at $z=0$ for each method as indicated. The box is 2$h^{-1}$Mpc on a side. The white circle indicates $M^{200}_{crit}$ for the halo, the black circle the same but for the G3-MUSIC simulation. }
\label{fig:Non-cooling_visual_dm_z0}
\end{figure*}

\begin{figure*}
\includegraphics[width=0.6\textwidth]{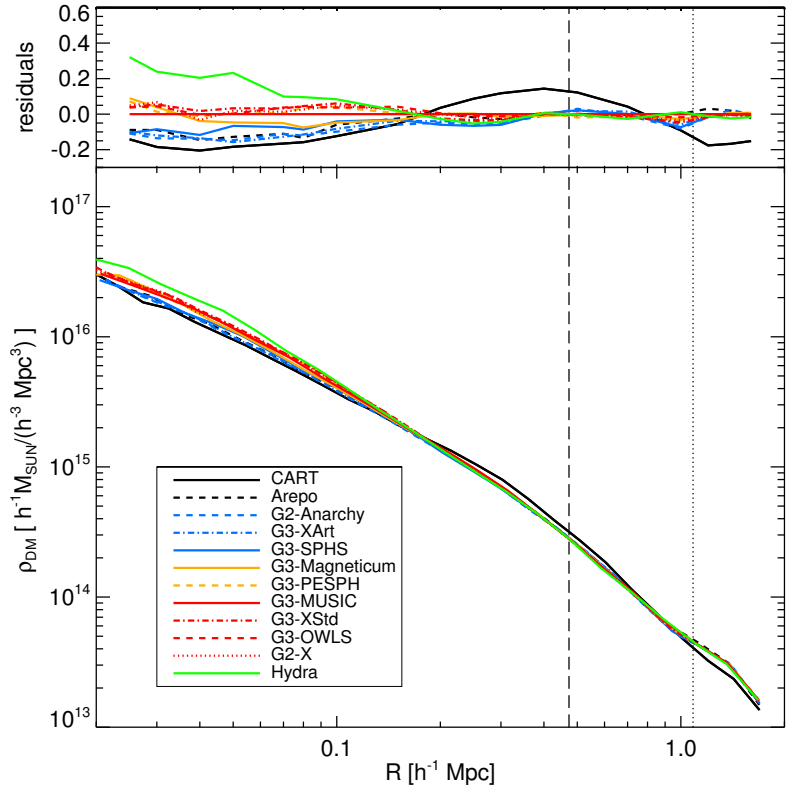}
\caption{Radial density profiles for the non radiative simulations at $z=0$ (bottom panel) and difference between the radial density profiles of each non radiative simulations at $z=0$ and the reference G3-MUSIC density profile (top panel). The vertical dashed line corresponds to $R_{2500}$ and the vertical dotted line to $R_{500}$ of the reference G3-MUSIC values.}
\label{fig:NR_DM_density_z0}
\end{figure*}

In order to perform a clean comparison of the various gas physics
implementations we carefully aligned the underlying
gravitational framework for each model. While
Figure~\ref{fig:DM_visual_z0} illustrates the range of outcomes that
result from a blind comparison using individual parameter choices, we can
choose a common parameter set for those quantities that control the
accuracy of the gravitational forces. For instance, for \gadget,
Table~\ref{tab:gadgetparams} gives the parameter choices made
independently by each group. The final row lists the common parameter set
adopted for the non-radiative comparison. For this common choice the
gravitational evolution of the nine \gadget\ simulations and \arepo\
is, as expected, essentially indistinguishable, as illustrated by
Figure~\ref{fig:Non-cooling_visual_dm_z0}.
Figure~\ref{fig:NR_DM_density_z0}  shows the radial density distribution and the difference relative to the G3-MUSIC simulation. For \hydra, the central gas density is so high that it steepens the central dark matter distribution relative to the other codes.

\label{lastpage}

\end{document}